\documentclass[12pt, a4paper]{article}
\usepackage[utf8]{inputenc} 
\usepackage{chet}
\usepackage{authblk}
\usepackage{slashed}
\usepackage{indentfirst}
\usepackage{amsmath}
\usepackage{amsfonts}
\usepackage{amssymb}
\usepackage{courier}
\usepackage{hyperref}
\newcommand{\m}[1]{\mathcal{#1}}

\newcommand{\B}[1]{\overline{#1}}

\newcommand{\WT}[1]{\widetilde{#1}}
\let\ampersand\&
\renewcommand*\&{and}

\usepackage[T1]{fontenc} 
\usepackage{xcolor}

\def \p{\partial}
\def \pb{\overline\partial}
\def \Jb{\overline J}

\def \a{\alpha}

\def \b{\beta}

\def \d{\delta}
\def \g{\gamma}

\def \L{\Lambda}

\def \O{\Omega}

\def \N{\nabla}
\def \Nb{\overline\nabla}

\def \r{\rho}

\def \s{\sigma}

\title{B-RNS-GSS heterotic string in curved backgrounds}
\author{Nathan Berkovits$^a$\footnote[1]{\texttt{nathan.berkovits@unesp.br}}, Osvaldo Chandia$^b$\footnote[2]{\texttt{ochandiaq@gmail.com}}, João Gomide$^{a}$\footnote[3]{\texttt{jpr.gomide@unesp.br}}, Lucas N. S. Martins$^a$\footnote[4]{\texttt{lucas\_nmartins@hotmail.com}}}
\affil{$^a$\textit{ICTP South American Institute for
Fundamental Research}\\
\textit{Instituto de Física Teórica, UNESP - Univ. Estadual Paulista}\\
\textit{Rua Dr. Bento Teobaldo Ferraz 271, 01140-070, São Paulo, SP, Brazil}\\
$^b$\textit{Departamento de Ciencias, Facultad de Artes Liberales}\\
\textit{Universidad Adolfo Ib\'a\~nez, Santiago, Chile. }}

\abstract{\quad The recently established B-RNS-GSS formalism is extended for the description of the heterotic superstring in curved backgrounds. We propose a generalized action and BRST charge defined in the small Hilbert space with the standard form of an $\m{N}=(1,0)$ worldsheet superconformal theory with superconformal generator $G$ and stress tensor $T$. We show that $\big\{G,G\big\}=-2T$ implies the D=10 N=1 supergravity and super-Yang-Mills equations of motion, as well as holomorphicity of the BRST charge.}

\begin{document} 
\maketitle
\flushbottom
\tableofcontents

\section{Introduction}

Although the Ramond-Neveu-Schwarz (RNS) formalism for the superstring has an elegant structure with $\m{N}=1$ worldsheet supersymmetry, the construction of vertex operators for spacetime fermions using this formalism is quite complicated. This makes it difficult to use the RNS formalism to compute scattering amplitudes involving external fermions or to describe Ramond-Ramond backgrounds \cite{Friedan:1985ge}.

After adding extra worldsheet variables to the RNS formalism which are spacetime spinors, it was recently
shown how to construct simple vertex operators for the spacetime fermions and make spacetime supersymmetry manifest while preserving N=1 worldsheet supersymmetry \cite{Berkovits:2013eqa, Berkovits:2016xnb, Berkovits:2021xwh}. This new formalism involves features of the RNS, Green-Schwarz-Siegel \cite{Siegel:1985xj} and pure spinor \cite{Berkovits:2000fe} descriptions of the superstring and was called the B-RNS-GSS formalism. 

An important open problem in superstring theory is to prove equivalence of the RNS and pure spinor
descriptions, and the B-RNS-GSS formalism will probably play a useful role as a bridge between the two descriptions. The pure spinor description has the advantage over the RNS description of manifest spacetime supersymmetry and can easily describe general D=10 supergravity backgrounds. As shown in \cite{Berkovits:2001ue},
nilpotence and holomorphicity at the classical level of the pure spinor BRST operator in curved heterotic and Type II supergravity backgrounds imply that the heterotic and Type II backgrounds obey the N=1 and N=2 D=10 supergravity equations of motion. 

In this paper, the B-RNS-GSS formalism will be defined in a curved heterotic supergravity background and nilpotence and holomorphicity of the BRST operator at the classical level will similarly be shown to imply the N=1 D=10 supergravity equations for the background. In the B-RNS-GSS formalism, the BRST operator has the standard
form of an $\m{N}=1$ worldsheet superconformal field theory, $Q= \int (c T + \gamma G + \gamma^2 b - b c \partial c)$,
and nilpotence and holomorphicity of $Q$ follow from requiring that $G$ and $T$ generate an $\m{N}=1$ worldsheet superconformal algebra. It should be possible to generalize these results to Type II supergravity backgrounds as well as to the computation of $\alpha'$ corrections, however, this will be left for later work.

In Section 2 of this paper, we review the B-RNS-GSS formalism in a flat background. In Section 3, we construct the worldsheet action and BRST operator in a curved heterotic supergravity background. In Section 4 we compute the constraints on the background superfields implied by requiring that $\big\{ G, G\big\}=-2T$. In Section 5 we show that these constraints imply that the BRST charge is holomorphic. In Section 6 we analyse the results from the previous sections. More precisely, we show that the constraints found in Section 4 are the D=10 N=1 supergravity and super-Yang-Mills equations of motion. In Section 7 we recast the problem in $\m{N}=(1,0)$ superfield language.  Finally, the appendices contain the conventions and formulas which will be used throughout the paper: Appendix \ref{SuperGeo} compiles the supergeometry conventions, whereas Appendix \ref{apa} and \ref{apb} compile the relevant equations for the $\big\{G,G\big\}=-2T$ and holomorphicity computations respectively.

\section{Review of the B-RNS-GSS formalism} \label{SectionBRNSGSS}

The first step to constructing the B-RNS-GSS formalism is to add the spacetime spinor variables 
$(\theta^\alpha, p_\alpha)$ and $(\Lambda^\alpha, w_\alpha)$ for $\alpha=1$ to 16 to the RNS worldsheet variables $(x^m, \psi^m,b,c, \beta, \gamma; \bar b, \bar c, \B{\rho}_{\mathcal{A}})$ with the worldsheet action
$$S = S_{RNS} + \int d^2 z (p_\alpha\bar\partial\theta^\alpha + w_\alpha \bar\partial \Lambda^\alpha)$$
and the left-moving BRST operator
$$Q = Q_{RNS} + \int dz (\Lambda^\alpha p_\alpha),$$
where $(\theta^\alpha, p_\alpha)$ and $(\Lambda^\alpha, w_\alpha)$ are left-moving fermionic and bosonic variables of conformal weight $(0,1)$. Using the usual quartet argument, the variables $(\theta^\alpha, p_\alpha, \Lambda^\alpha, w_\alpha)$ decouple from the BRST cohomology so the physical superstring 
spectrum is unchanged.

The second step is to perform a similarity transformation on all worldsheet variables as
${\cal O} \to e^{-R} {\cal O} e^{R}$ where 
\begin{equation} \label{simR}
R = \oint dz \Big[ {{(\Lambda\gamma_m\theta)}\over {2\gamma}}\psi^m + c w_\alpha \partial \theta^\alpha\Big]. 
\end{equation}
After the similarity transformation, the worldsheet action and BRST operator have the manifestly spacetime supersymmetric
form 
\begin{equation} \label{SBRNSGSS}
    \begin{split}
    S_{R-RNS-GSS}=\int d^2z\Big(\frac{1}{2}\Pi_m\overline \Pi^{m}+B_{het}+\frac{1}{2}\psi^m\B{\partial}\psi_m+d_{\alpha}\B{\partial}\theta^{\alpha}+w_{\alpha}\B{\partial}\Lambda^{\alpha}+\\+b\B{\partial}c+\beta\B{\partial}\gamma+\B{b}\partial\B{c}+\frac{1}{2}\overline\rho^{\mathcal{A}}\partial\overline\rho_{\mathcal{A}}\Big),
    \end{split}
\end{equation}
\begin{equation} \label{Qflat}
\begin{split}
    Q_{B-RNS-GSS}=\oint dz\Big(&cT_{B-RNS-GSS}-bc\partial c+\gamma^2b+\Lambda^{\alpha}d_{\alpha}+\gamma\psi^m \Pi_m+\gamma^2\partial\theta^{\alpha}w_{\alpha}\\
    &+\frac{1}{2\gamma}\big(\Lambda\gamma^m\Lambda\big)\psi_m \Big),
\end{split}
\end{equation}
where $\B{\rho}_{\m{A}}$ are the right-moving fermionic variables of conformal weight ${1\over 2}$ in the fundamental representation of $E_8\times E_8$ or $SO(32)$,
\begin{equation}
B_{het}\equiv\frac{1}{4}(\overline\partial \theta^{\alpha}\Pi^m -\overline \Pi^m \partial\theta^{\alpha})(\theta\gamma_{m})_{\alpha}\,,    
\end{equation}
\begin{equation}
    \Pi^m\equiv\partial x^m-\frac{1}{2}(\partial\theta\gamma^m\theta),\quad 
   \overline \Pi^m\equiv\overline\partial x^m-\frac{1}{2}(\overline\partial\theta\gamma^m\theta),\quad d_{\alpha}\equiv p_{\alpha}-\frac{1}{2}\Big(\partial x^m-\frac{1}{4}\partial\theta\gamma^m\theta\Big)\big(\gamma_m\theta\big)_{\alpha}
\end{equation}
are defined as in the Green-Schwarz-Siegel formalism, and the stress-energy tensor is 
\begin{equation}\label{stress}
    T_{B-RNS-GSS}=-\frac{1}{2}\Pi^m \Pi_m-d_{\alpha}\partial\theta^{\alpha}-\frac{1}{2}\psi^m\partial\psi_m-w_{\alpha}\partial\Lambda^{\alpha}-\beta\partial\gamma-\frac{1}{2}\partial\big(\beta\gamma\big)-b\partial c-\partial\big(bc\big).
\end{equation}

Since the similarity transformation $R$ of (\ref{simR}) involves inverse powers of $\gamma$,
it is only well-defined in the large Hilbert space including the $\xi$ zero mode where 
$\gamma = \eta e^{\phi}$, $\beta = \partial \xi e^{-\phi}$ and $\gamma^{-1} = \xi e^{-\phi}$. As discussed in \cite{Berkovits:2021xwh,Berkovits:2001us,Kroyter:2009rn}, one method for working in the large Hilbert space is to generalize
the usual BRST operator $Q_{small}$ to $Q_{large} \equiv Q_{small} - \oint dz \eta$ and define 
physical states to be in the cohomology of $Q_{large}$. In terms of $Q_{large}$, picture-changing is a BRST-trivial operation since $Q_{large} (\xi V) = Q_{small}(\xi) V - V$ when $V$ is annihilated by $Q_{large}$. 

However, this method in the large Hilbert space breaks the $\m{N}=1$ worldsheet supersymmetry, and it will be more convenient in this paper to preserve the $\m{N}=1$ worldsheet supersymmetry by working in the small Hilbert space. This can be accomplished by ``untwisting'' 
\begin{equation}\label{untwist}
\Lambda^\alpha \to \gamma \Lambda^\alpha, \quad w_\alpha \to {1\over\gamma} w_\alpha
\end{equation}
so that the untwisted $(\Lambda^\alpha, w_\alpha)$ carry conformal weight $({1\over 2}, {1\over 2})$\footnote{ This untwisting procedure shifts the conformal anomaly contribution of the $(\Lambda^\alpha, w_\alpha)$ variables from $+32$ to $-16$, so the untwisted BRST operator is no longer nilpotent. To make this untwisting procedure consistent at the
quantum level, one needs to add to the RNS variables not only the $(\theta^\alpha, p_\alpha, \Lambda^\alpha, w_\alpha)$ variables, but also the non-minimal variables $(\overline\Lambda_\alpha, \overline w^\alpha, R_\alpha, S^\alpha)$ where $(\overline\Lambda_\alpha, 
\overline w^\alpha)$ and $(R_\alpha, S^\alpha)$ are 32 left-moving bosons and fermions of conformal weight $(0,1)$ \cite{Berkovits:2022}. The non-minimal BRST operator is obtained by starting with 
$Q = Q_{RNS} + \int dz (\Lambda^\alpha p_\alpha + \overline w^\alpha R_\alpha)$ and performing
a similarity transformation with 
$$R =\oint dz \Big[ {{(\Lambda\gamma_m\theta)}\over {2\gamma}} \psi^m + c (w_\alpha \partial \theta^\alpha + S^\alpha \partial \overline\Lambda_\alpha)\Big].$$
So $(\overline\Lambda_\alpha, \overline w^\alpha, R_\alpha, S^\alpha)$ play a similar role to the non-minimal variables $(\overline\lambda_\alpha, \overline w^\alpha, r_\alpha, s^\alpha)$ in the pure spinor
formalism.

To obtain a nilpotent BRST operator after untwisting, one can now define $J = \xi\eta - \Lambda^\alpha w_\alpha + R_\alpha S^\alpha$ and untwist
\begin{equation}
\Lambda^\alpha \to \gamma \Lambda^\alpha, \quad w_\alpha \to {1\over\gamma} w_\alpha,\quad
R_\alpha \to \gamma R_\alpha, \quad S^\alpha \to {1\over\gamma} S^\alpha,
\end{equation}
so that the change in the conformal anomaly of $(\Lambda^\alpha, w_\alpha)$ is cancelled
by the change in the conformal anomaly of $(R_\alpha, S^\alpha)$. However, as in the pure 
spinor formalism, the non-minimal variables $(\bar\Lambda_\alpha, \bar w^\alpha, R_\alpha, S^\alpha)$ are not needed for constructing massless vertex operators in the B-RNS-GSS formalism and will be ignored for the curved supergravity backgrounds discussed in this paper.
}.
This untwist is accompanied by a shift $\beta \to \beta - \gamma^{-1} \Lambda^\alpha w_\alpha$, and
the resulting BRST operator has the standard $\m{N}=1$ worldsheet supersymmetric form
\begin{equation} \label{standard}
Q_B\equiv\oint dz ~j_B = \oint dz \left( c T  - bc\p c + b\g^2 + \gamma G \right),
\end{equation}
where 
\begin{equation} \label{Guntwist}
G = \Lambda^{\alpha}d_{\alpha}+\psi^m \Pi_m+\partial\theta^{\alpha}w_{\alpha}+\frac{1}{2}\big(\Lambda\gamma^m\Lambda\big)\psi_m,
\end{equation}
$T = T_{B-RNS-GSS} + \frac12 \partial (\Lambda^\alpha w_\alpha)$ is the untwisted stress-tensor, and $T_{B-RNS-GSS}$ is defined
in (\ref{stress}).

In terms of the untwisted variables of (\ref{untwist}), physical vertex operators need to not only be in the cohomology of the $\m{N}=1$ BRST operator of (\ref{Qflat}), they need to also
be in the small Hilbert space after twisting the variables $\Lambda^\alpha \to \gamma^{-1} \Lambda^\alpha$ and $w_\alpha \to \gamma w_\alpha$. Defining the $U(1)$ generator 
\begin{equation}
J = \xi\eta -\Lambda^\alpha w_\alpha
\end{equation}
so that $\Lambda^\alpha$ and $\beta$ carry charge $+1$ and $w_\alpha$ and $\gamma$ carry charge $-1$, this
requires that all terms in the vertex operator carry non-positive charge with respect to $J$. Note that a similar
requirement was imposed on vertex operators in the untwisted pure spinor formalism of \cite{Berkovits:2016xnb}.

To construct the worldsheet action in a curved supergravity background, it will be useful to start by constructing the massless super-Yang-Mills and supergravity vertex operators corresponding to the linearized deformation around a flat
background \cite{Witten:1985nt}.
As discussed in \cite{Berkovits:2021xwh,Berkovits:2013eqa}, the massless integrated vertex operator in the B-RNS-GSS formalism for the open superstring is:
\begin{equation}
    \int dz~U_{open}= \int dz G (-\Lambda^{\alpha}A_{\alpha}(X,\theta)+ \psi^m A_m(X,\theta)+w_{\alpha}W^{\alpha}(X,\theta))
\end{equation}
\begin{equation}\label{open}
\begin{split}
   = \int dz\Bigg(&\partial\theta^{\alpha}A_{\alpha}(X,\theta)+\Pi^m A_m(X,\theta)+d_{\alpha}W^{\alpha}(X,\theta)\\
    &-\frac{1}{2}\Big(\psi^m\psi^n-\frac{1}{2}\Lambda\gamma^{mn}w\Big)F_{mn}(X,\theta)-\psi^m w_{\alpha}\partial_m W^{\alpha}(X,\theta)\Bigg),
\end{split}
\end{equation} 
where $G(A)$ denotes the simple pole of $G$ with $A$. Note that all terms in (\ref{open}) contain zero or $-1$ U(1) charge, so the vertex operator is in the small Hilbert space when expressed in terms of the
twisted variables. 
The heterotic string massless integrated vertex operators can be easily constructed from the product of the open vertex operator with $\int d\B{z}\B{\partial}x^m$ or $\int d\B{z}\B{J}^I$ for the supergravity and super-Yang-Mills cases, respectively. Consequently, we end up with
\begin{equation} \label{SUGRAvertex}
    \int d^2z~U_{SUGRA}=\int d^2z\Big[\partial\theta^{\alpha}A_{\alpha m}+\Pi^n A_{nm}+d_{\alpha}E_m\,^{\alpha}+\frac{1}{2}N_{np}\Omega_{m}\,^{np}-\psi^n w_{\alpha}\partial_nE_m\,^{\alpha}\Big]\B{\partial}x^m,
\end{equation}
\begin{equation} \label{SYMvertex}
    \int d^2z~U_{sYM}=\int d^2z\Big[\partial\theta^{\alpha}A_{\alpha I}+\Pi^nA_{nI}+d_{\alpha}W_I\,^{\alpha}+\frac{1}{2}N_{np}U_I^{np}-\psi^nw_{\alpha}\partial_nW_I\,^{\alpha}\Big]\B{J}^I,
\end{equation}
where we denote
\begin{equation}
    N_{np}\equiv-\psi_n\psi_p+\frac{1}{2}\Lambda\gamma_{np}w.
\end{equation} 
The requirement of BRST-invariance on these vertices implies the linearized SUGRA/SYM equations of motion for the $\big\{A_{\alpha m},A_{mn},E_m\,^{\alpha},\Omega_m\,^{np},A_{\alpha I},A_{aI},W_I\,^{\alpha},U_I\,^{ab} \big\}$ superfields:
\begin{equation}
D_{(\alpha}A_{\beta)m}=\gamma^n_{\alpha\beta}A_{nm},\qquad D_{(\alpha}A_{\beta)I}=\gamma^m_{\alpha\beta}A_{mI},
\end{equation}
\begin{equation}
    D_{\alpha}A_{mn}-\partial_{m}A_{\alpha n}=\gamma_{m\alpha\beta}E_n\,^{\beta},\qquad D_{\alpha}A_{mI}-\partial_{m}A_{\alpha I}=\gamma_{m\alpha\beta}W_I\,^{\beta},
\end{equation}
\begin{equation}
    \partial_{[m}A_{n]p}=\Omega_{pmn},\qquad \partial_{[m}A_{n]I}=U_{mn I},
\end{equation}
\begin{equation}
    D_{\alpha}E_m\,^{\beta}=\Omega_{m\alpha}\,^{\beta},\qquad D_{\alpha}W_I\,^{\beta}=U_{I\alpha}\,^{\beta},
\end{equation}
\begin{equation}
    \gamma^n_{\alpha\beta}\partial_nE_m\,^{\beta}=0\,,\qquad \gamma^m_{\alpha\beta}\partial_mW_I\,^{\beta}=0.
\end{equation}
Note that this is the same set of background superfields and linearized equations of motion as the one we encounter in the pure spinor formalism \cite{Berkovits:2001ue,Berkovits:2000fe}.

\section{B-RNS-GSS heterotic string in curved backgrounds}

We will now write the expressions for the most general worldsheet action and BRST charge in the small Hilbert space as defined in the last section. In other words, after twisting
$\Lambda^\alpha \to \gamma^{-1} \Lambda^\alpha$ and $w_\alpha \to \gamma w_\alpha$, all terms
in the action and all terms except for $ \frac{\gamma}{2}(\Lambda\gamma^m \Lambda)\psi_m$ in the BRST charge must have non-negative powers of $\gamma$. The most general action for this system can be written as
\begin{equation} \label{Sdyn}
\begin{split}
    S=\frac{1}{2\pi\alpha'}\int d^2z\Big[&\frac{1}{2} E_a\B{ E}^a+\frac{1}{2} E^A\B{ E}^BB_{BA}+\frac{1}{2}\psi_a\B{\nabla}\psi^a+d_{\alpha}\B{ E}^{\alpha}+w_{\alpha}\B{\nabla}\Lambda^{\alpha}+b\B{\partial}c+\beta\B{\partial}\gamma+\B{b}\partial\B{c}\\
    &+\frac{1}{2}\B{\rho}_{\m{A}}\nabla\B{\rho}_{\m{A}}+d_{\alpha}\B{J}^IW_I\,^{\alpha}+\frac{1}{2}\psi^a\psi^b\B{J}^IU_{Iab}+\Lambda^{\alpha}w_{\beta}\B{J}^IU_{I\alpha}\,^{\beta}\\
    &+\psi^aw_{\alpha}\Big(\B{ E}^AC_{Aa}\,^{\alpha}+\B{J}^IC_{Ia}\,^{\alpha}\Big)+w_{\alpha}w_{\beta}\Big(\B{ E}^AY_A\,^{\alpha\beta}+\B{J}^IY_I\,^{\alpha\beta}\Big)\Big].
\end{split}
\end{equation}
The covariant derivatives are defined as
\begin{align}
&\Nb\psi^a =\pb\psi^a + \psi^b (\pb Z^M \O_{Mb}{}^a) ,\cr
&\Nb\L^\a = \pb\L^\a + \L^\b (\pb Z^M \O_{M\b}{}^\a) ,\cr
&\N\rho_{\cal A} = \p\rho_{\cal A} + (t^I_{\cal AB} \p Z^M A_{MI}) \rho_{\cal B},
\label{nablas}
\end{align}
and the connection $\O_{M\a}{}^\b$ has the general form
\begin{align}
\O_{M\a}{}^\b = \O_M\d_\a^\b + \frac14 \O_{Mab}(\g^{ab})_\a{}^\b  + \O_{Mabcd} (\g^{abcd})_\a{}^\b,
\label{Ogen}
\end{align} 
where $\O_M$ is a scale connection and the presence of $\O_{Mabcd}$ is allowed since $\L^\a$ is unconstrained. 
Similarly, the background field $U_{I\a}{}^\b$ can be expanded as
\begin{align}
U_{I\a}{}^\b =  U_I\d_\a^\b + \frac14 U_{Iab}(\g^{ab})_\a{}^\b + U_{Iabcd} (\g^{abcd})_\a{}^\b .
\label{Ugen}
\end{align}
The background field $A_{MI}$ is the Yang-Mills connection for the gauge group, $t^I_{\cal AB}$ are the generators of the Lie algebra for the gauge group and $\Jb^I\equiv\frac12 t^I_{\cal AB}\overline\rho_{\cal A} \overline\rho_{\cal B}$. Finally, we stress the fact that the background superfields $\big\{B, E, \O,  A, W, U, C, Y\big\}$ are functions of the target space coordinates $Z^M$.

The BRST charge retains the standard $\m{N}=(1,0)$ worldsheet supersymmetric form
\begin{align}
Q_B\equiv\oint d\s~ j_B = \oint d\s \Bigg( cT - c\Big(\frac{3}{2}\beta\partial\gamma+\frac{1}{2}\gamma\partial\beta\Big) + bc\p c + b\g^2 + \gamma G\Bigg),
\label{BRST}
\end{align}
but now we generalize
\begin{equation} \label{Ggeneral2}
\begin{split}
    G=&\ \frac{1}{2}\Lambda^{\alpha}\Lambda^{\beta}\psi_c\big(\gamma^c\big)_{\alpha\beta}\\
    &+\Lambda^{\alpha}d_{\alpha}+\frac{1}{2}\Lambda^{\alpha}\Lambda^{\beta}w_{\gamma}G_{\alpha\beta}\,^{\gamma}+\frac{1}{2}\Lambda^{\alpha}\psi^b\psi^cG_{\alpha bc}\\
    &+\psi^a E_a+\frac{1}{6}\psi^a\psi^b\psi^cG_{abc}+\Lambda^{\alpha}\psi^bw_{\gamma}G_{\alpha b}\,^{\gamma}\\
    &+w_{\alpha} E^{\alpha}+\frac{1}{2}\psi^a\psi^bw_{\gamma}G_{ab}\,^{\gamma}+\frac{1}{2}\Lambda^{\alpha}w_{\beta}w_{\gamma}G_{\alpha}\,^{\beta\gamma}\\
    &+\frac{1}{2}\psi^aw_{\beta}w_{\gamma}G_a\,^{\beta\gamma}+\frac{1}{6}w_{\alpha}w_{\beta}w_{\gamma}G^{\alpha\beta\gamma}
\end{split}
\end{equation}
and
\begin{align}
T= &-\frac12 E_a E^a - d_\a E^\a - \frac12 \psi_a \N \psi^a -\frac12 w_\a \N \L^\a  + \frac12 \L^\a \N w_\a \cr
&- \psi^a w_\a E^A C_{Aa}{}^\a -  w_\a w_\b E^A Y_A{}^{\a\b}.
\label{T}
\end{align}
Of course, the background fields denoted by $G$ in (\ref{Ggeneral2}) are also functions of the target space coordinates $Z^M$.

As mentioned previously, the background fields which were introduced above parametrize the deformation away from the flat background. The physical deformations are the ones such that
\begin{equation}
    \big\{ G, G\big\}=-2T.
\end{equation}
Therefore, in order to find them, we will now require that the most general small Hilbert space expressions for $S$ and $G$ written above are such that $G$ squares to $-2T$. This procedure imposes constraints on the background fields. As we will see, these imply that $G$ is holomorphic and will turn out to be the D=10 N=1 supergravity and super-Yang-Mills constraints of \cite{Berkovits:2001ue}.

\section{$\big\{ G, G\big\}=-2T$ condition} \label{SectionG2}
We start by listing the canonical commutation relations that will be needed to compute $\big\{ G, G\big\}=-2T  $. These are
\begin{equation}
    \big[\psi^a(\sigma),\psi^b(\sigma')\big]=\eta^{ab}\delta(\sigma-\sigma'),\qquad \big[\Lambda^{\alpha}(\sigma),w_{\beta}(\sigma')\big]=\delta^{\alpha}_{\beta}\delta(\sigma-\sigma'),
    \label{com1}
\end{equation}
\begin{equation}
    \big[c(\sigma),b(\sigma')\big]=\delta(\sigma-\sigma'),\qquad\big[\gamma(\sigma),\beta(\sigma')\big]=\delta(\sigma-\sigma'),
    \label{com2}
\end{equation}
\begin{equation}
    \big[\B{J}^I(\sigma),\B{J}^J(\sigma')\big]=f_K\,^{IJ}\B{J}^K(\sigma)\delta(\sigma-\sigma').
    \label{com3}
\end{equation}
For the superspace coordinates, the conjugate momentum $P_M$ is defined by
\begin{equation}
     (-1)^M P_M \equiv
     \frac{\d S}{\d(\p_\tau Z^M)},\qquad [Z^M(\s) , P_N(\s') ] = \d^M_N \d(\s-\s') ,
\label{PZ}
\end{equation}
where $\p_\tau=\frac12(\pb+\p)$ (and $\p_\s=\frac12(\pb-\p)$). It follows that $P_M$ is written as
\begin{equation}
\begin{split}
    (-1)^MP_M=&\partial_{\tau}Z^NG_{NM}+\partial_{\sigma}Z^NB_{NM}-E_M\,^{\alpha}d_{\alpha}-\frac{1}{2}\Omega_{Mab}\psi^a\psi^b+\Omega_{M\alpha}\,^{\beta}\Lambda^{\alpha}w_{\beta}+A_{MI}\B{J}^I\\
    &+(-1)^M\psi^aw_{\alpha}C_{Ma}\,^{\alpha}+w_{\alpha}w_{\beta}Y_M\,^{\alpha\beta}.
\end{split}
\end{equation}
This expression can be inverted to yield
\begin{align}
d_\a &= (-1)^{M+1} E_\a{}^M P_M + \p_\s Z^M B_{M\a} - \frac12 \O_{\a ab} \psi^a \psi^b + \O_{\a\b}{}^\g \L^\b w_\g + A_{\a I} \Jb^I \cr
&\qquad- \psi^a w_\b C_{\a a}{}^\b + w_\b w_\g Y_\a{}^{\b\g}  
\label{dis}
\end{align}
and, similarly,
\begin{align}
d_a\equiv\p_\tau Z^M E_{Ma} &= (-1)^M E_a{}^M P_M - \p_\s Z^M B_{Ma} + \frac12 \O_{abc} \psi^b \psi^c - \O_{a\a}{}^\b \L^\a w_\b - A_{aI} \Jb^I \cr
&\qquad-  \psi^b w_\a C_{ab}{}^\a - w_\a w_\b Y_a{}^{\a\b}.  
\label{p0Z}
\end{align}
It follows that
\begin{gather}
    E_a=d_a-\partial_{\sigma}Z^ME_{Ma},\\
    \B{E}_a=d_a+\partial_{\sigma}Z^ME_{Ma},\\
    \B{E}^{\alpha}=-\B{J}^IW_I\,^{\alpha},\\
    E^{\alpha}=-\B{J}^IW_I\,^{\alpha}-2\partial_{\sigma}Z^ME_{M}\,^{\alpha}.
\end{gather}

It is also important to define explicitly what we mean by the $\big\{ G, G\big\}$ anti-commutator:
\begin{equation} \label{QQdef}
    \big\{ G, G\big\}=\oint\frac{d\sigma}{2 \pi i}\oint\frac{d\sigma'}{2 \pi i}\big\{ G(\sigma), G(\sigma')\big\}.
\end{equation}

One final comment is in order before we start the computation. The action and BRST charge are invariant under the gauge transformations
\begin{equation} \label{red1}
    \delta d_{\alpha}=-\Lambda^{\beta}w_{\gamma}\omega_{\alpha\beta}\,^{\gamma},\qquad \delta G_{\alpha\beta}\,^{\gamma}=\omega_{(\alpha\beta)}\,^{\gamma},
\end{equation}
\begin{equation} \label{red2}
    \delta\Omega_{\alpha\beta}\,^{\gamma}=-\omega_{\alpha\beta}\,^{\gamma}, \qquad \delta U_{I\alpha}\,^{\beta}=-W_I\,^{\rho}\omega_{\rho\alpha}\,^{\beta},
\end{equation}
which can be used to gauge the $\Omega_{\alpha}\,^{bcde}$ background field to zero. Note that after this gauge-fixing, the gauge symmetries parameterized by
\begin{equation} \label{unfixed}
    \omega_{\alpha\beta}\,^{\gamma}=\omega_{\alpha}\delta_{\beta}^{\gamma}+\frac{1}{4}\omega_{\alpha}\,^{ab}\big(\gamma_{ab}\big)_{\beta}\,^{\gamma}
\end{equation}
remain unfixed. Then, consider the decomposition of $G_{\alpha\beta}\,^{\gamma}$ into irreducible representations
\begin{equation}
    G_{\alpha\beta}\,^{\gamma}=G_a\,^{\gamma}\big(\gamma^a\big)_{\alpha\beta}+F_{abcde}\,^{\gamma}\big(\gamma^{abcde}\big)_{\alpha\beta},
\end{equation}
\begin{equation}
    G_a\,^{\gamma}=\widetilde{G}^{144}_a\,^{\gamma}+G^{16}_{\delta}\big(\gamma_a\big)\,^{\delta\gamma},
\end{equation}
\begin{equation}
    F_{abcde}\,^{\gamma}=\widetilde{F}^{1440}_{[abcd|\delta|}\big(\gamma_{e]}\big)^{\delta\gamma}+\widetilde{F}^{560}_{[ab|\delta|}\big(\gamma_{cde]}\big)^{\delta\gamma}+F^{16}_{\delta}\big(\gamma_{abcde}\big)^{\delta\gamma},
\end{equation}
where the superscripts denote the dimension of the irrep and the tildes over the tensors denote that they are $\gamma$-traceless. We see that the $2176$ components of $G_{\alpha\beta}\,^{\gamma}$ split up in the following way
\begin{equation}
    2176=16\oplus 16\oplus 144\oplus 560 \oplus 1440.
\end{equation}
Now, consider also the decomposition of some field
\begin{equation}
    \Phi_{\alpha\beta}\,^{\gamma}=\Phi_{\alpha}\delta^{\gamma}_{\beta}+\frac{1}{4}\Phi_{\alpha}\,^{ab}\big(\gamma_{ab}\big)_{\beta}\,^{\gamma},
\end{equation}
 into irreducible representations
\begin{equation} \label{Weylirreps}
    \Phi_{\alpha}=\Phi^{16}_{\alpha},
\end{equation}
\begin{equation} \label{Lorentzirreps}
    \Phi_{\alpha}\,^{ab}=\WT{L}^{560}_{\alpha}\,^{ab}+\WT{L}^{144 \beta[a}\big(\gamma^{b]}\big)_{\beta\alpha}+L^{16}_{\beta}\big(\gamma^{ab}\big)_{\alpha}\,^{\beta},
\end{equation}
from which we conclude that its 736 components split up into
\begin{equation}
    736=16\oplus16\oplus144\oplus560.
\end{equation}
Note that none of the irreps in (\ref{Weylirreps}) and (\ref{Lorentzirreps}) are in the kernel of the linear map
\begin{equation}
    \Phi_{\alpha\beta}\,^{\gamma}\to\Phi_{(\alpha\beta)}\,^{\gamma}.
\end{equation}
Thus, we can also use the leftover gauge symmetry of (\ref{unfixed}) to gauge $\big(G^{16}_{\alpha},F^{16}_{\alpha},\WT{F}^{144\alpha a},\WT{F}^{560}_{\alpha ab}\big)$ to zero. However, note that $\WT{F}^{1440}_{abcd\delta}$ stays intact after this procedure, and consequently we end up with
\begin{equation} \label{red4}
    G_{\alpha\beta}\,^{\gamma}=\widetilde{F}^{1440}_{abcd\delta}\big(\gamma_e\big)\,^{\delta\gamma}\big(\gamma^{abcde}\big)_{\alpha\beta}\equiv G^{1440}_{\alpha\beta}\,^{\gamma}.
\end{equation}

\subsection{$\big\{ G, G\big\}=-2T$ computation}
Now we compute
\begin{align}
\left( \oint  G\right)  G = \oint d\s' \big\{ G(\s') ,  G (\s)\big\} .
\label{q12}
\end{align}
To organize this calculation, we separate the contributions according to their U(1) charge. As we will see, (\ref{q12}) has the form
\begin{align}
\left( \oint  G\right)  G = \sum_{n=-4}^{4} {\cal G}_n,
\label{gGgG}
\end{align}
where the term $\m{G}_n$ has U(1) charge $n$. In general, the contributions will either give constraints for the supergravity and super Yang-Mills superfields (torsions, H-fluxes and Yang-Mills curvatures), will determine the background superfields in the action and in $G$ in terms of those, or will be equivalent to linear combinations of the Bianchi identities from (\ref{bianchiCOMP}) which vanish identically. For the particular cases of ${\cal G}_0,\m{G}_{-1}$ and $\m{G}_{-2}$, there are also contributions which will sum to $-2T$.

Before we begin, we refer the reader to Appendix \ref{SuperGeo} for our conventions related to supergeometry. In particular, we use
\begin{equation}
    T_{NM}{}^A=\partial_{[N}E_{M]}\,^A+(-1)^{N(B+M)}E_M\,^B\Omega_{NB}\,^A-(-1)^{MB}E_N\,^B\Omega_{MB}\,^A,
\end{equation}
\begin{equation}
    T_{BC}\,^A=(-1)^{B(M+C)}E_C\,^ME_B\,^NT_{NM}\,^A=f_{BC}\,^A+\Omega_{[BC]}\,^A,
\end{equation}
or equivalently
\begin{equation}
\nabla_{A}=E_{A}\,^{M}\partial_{M}-\Omega_{AB}\,^{C}\Sigma_{C}\,^{B}\,,\qquad [\nabla_{A},\nabla_{B}]=-T_{AB}\,^{C}\nabla_{C}-R_{ABC}\,^{D}\Sigma_{D}\,^{C}.    
\end{equation}
We also refer the reader to Appendix \ref{apa} where the relevant commutators for the computation are organized.

Using the commutators shown in (\ref{com1})-(\ref{com3}), ${\cal G}_4$ is given by
\begin{align}
{\cal G}_4=\frac14 (\L\g^a\L) (\L\g_a\L) 
\label{G4}
\end{align}
which vanishes as a consequence of the Fierz identity  $\g^a_{\rho(\a} \g^b_{\b\g)}\eta_{ab}=0$. 

The contribution with $U(1)$-charge +3 is given by
\begin{align}
{\cal G}_3=\frac13\psi^a \L^\a \L^\b \L^\g \left( 2\O_{(\a} (\g_a)_{\b\g)} + (\g_a)_{\r(\a} G_{\b\g)}{}^\r + (\g^b)_{(\a\b} G_{\g)ba} \right).
\label{G3}
\end{align}
It is going to be shown below that this expression is zero after determining the value of the $G$'s in terms of torsion components. 

The contribution with $U(1)$-charge +2 is given by
\begin{align}
{\cal G}_2 =& \frac12 \L^\a \L^\b \B{E}^a \left( T_{\a\b a}-H_{\a\b a} \right) + \frac12 \L^\a \L^\b E^\g H_{\a\b\g}+ \L^\a \L^\b \Jb^I \left( F_{\a\b I}+\frac12 W_I{}^\g H_{\g\a\b} \right)  \cr
&+ \L^\a \L^\b E^a \left( \frac12\left( T_{\a\b a} + H_{\a\b a} \right) + (\g_a)_{\a\b} \right)+  \L^\a \L^\g d_\g \left( G_{\a\b}{}^\g + T_{\a\b}{}^\g \right)   \cr
&+\L^\a \L^\b \L^\g w_\r \left( \g^a_{\a\b} \left( G_{a\g}{}^\r - C_{\g a}{}^\r \right) + R_{\a\b\g}{}^\r + \N_\a G_{\b\g}{}^\r - G_{\a\b}{}^\s G_{\s\g}{}^\r \right) \cr
&+\frac12 \psi^a \psi^b \L^\a \L^\b \left( \N_{(\a} G_{\b)ab} + \g^c_{\a\b} G_{cab} - G_{\a\b}{}^\g G_{\g ab} + G_{(\a a}{}^c G_{\b)cb} - R_{\a\b ab} \right) \cr 
&+2\psi^a\psi^b \L^\a (\g_a\L)_\b \left( G_{b\a}{}^\b - C_{\a b}{}^\b - \O_b \d_\a^\b -\O_b{}^{cdef}(\g_{cdef})_\a{}^\b  \right) .
\label{G2}
\end{align}
The vanishing of the first line leads to the constraints
\begin{align}
T_{\a\b a}-H_{\a\b a}=0,\quad H_{\a\b\g}=0,\quad F_{\a\b I}=-\frac{1}{2}W_I\,^{\gamma}H_{\gamma\alpha\beta}=0 .
\label{cc1}
\end{align}
The vanishing of the first term of the second line gives the constraint
\begin{align}
T_{\a\b a}=-(\g_a)_{\a\b} .
\label{T=gama}
\end{align}
Note that in Equations (\ref{cc1}) and (\ref{T=gama}) we already start to obtain some of the D=10 N=1 SUGRA/SYM constraints of \cite{Berkovits:2001ue}. The vanishing of the term with $d_{\alpha}$ determines $G_{\alpha\beta}\,^{\gamma}$ as
\begin{align}
G_{\a\b}{}^\g= -T_{\a\b}{}^\g ,
\label{X1is}
\end{align}
where we recall that only the 1440 irrep is present. The vanishing of the third line, together with the Bianchi identity for $R_{(\alpha\beta\gamma)}\,^{\delta}$ and Equation (\ref{X1is}), imply that $G_{\alpha a}\,^{\beta}$ satisfies
\begin{align}
G_{a\a}{}^\b-C_{\a a}{}^\b - T_{a\a}{}^\b =0.
\label{}
\end{align}
Note that $\frac12 G_{\a(ab)}$ does not appear in (\ref{G2}). In fact, in the term $G_{(\a a}{}^c G_{\b)cb}$ the antisymmetric combination  $\frac12 G_{\a[ab]}$ survives after multiplying by $\psi^a\psi^b\L^\a\L^\b$. Therefore, the use of the Bianchi identity  for $R_{[\alpha\beta c]}\,^d$ allows to determine
\begin{align}
G_{\a ab}=\frac12 T_{\a[ab]},\quad G_{cab}=-\frac12 T_{c[ab]}.
\label{x2x4}
\end{align}
Combining these results with the last line of (\ref{G2}), one obtains that $T_{a\a}{}^\b$ satisfies the constraint
\begin{align}
T_{a\a}{}^\b = \d_\a^\b \O_a +  (\g_{bcde})_\a{}^\b \O_a{}^{bcde} .
\label{cc5}
\end{align}
As it was mentioned above, now we can see that the constraints in Equations (\ref{X1is}) and (\ref{x2x4}) imply that ${\cal G}_3$ is simply the Bianchi identity involving $\N_{(\a} T_{\b\g)}{}^a$, and therefore 
\begin{align}
{\cal G}_3 = 0.
\label{G3is0}
\end{align}

Consider now ${\cal G}_1$. Using the constraints obtained so far, ${\cal G}_1$ is given by 
\begin{align}
&{\cal G}_1 = -\psi^a\L^\a\B{E}^b\left( H_{\a ab} - T_{\a (ab)} \right) + \psi^a \L^\a E^b H_{\a ab} - 2 \psi^a \L^\a \Jb^I \left( F_{a\a I} + (\g_a)_{\a\b} W_I{}^\b \right) \cr
&+\psi^a \L^\a\L^\b w_\g \left( \N_a T_{\a\b}{}^\g + \N_{(\a} T_{\b)a}{}^\g +T_{a(\a}{}^A T_{A\b)}{}^\g + T_{\a\b}{}^\r T_{\r a}{}^\g - R_{a(\a\b)}{}^\g \right) \cr
&+2\psi^a \L^\a (\g_a\L)_\b w_\g \left( 2Y_\a{}^{\b\g} + G_\a{}^{\b\g} \right) + \psi^a (\L\g^b\L) w_\a \left( C_{[ab]}{}^\a - G_{ab}{}^\a \right) \cr
&+\L^\a \psi^a \psi^b \psi^c \left( \frac13 \N_\a T_{abc} - \N_a T_{\a bc} + T_{\a a}{}^d T_{dbc} - T_{\a\a}{}^\b T_{\b cb} - R_{\a abc} + (\g_a)_{\a\b} \left( G_{bc}{}^\b - C_{[bc]}{}^\b \right) \right) . \cr
&
\label{G1}
\end{align}
The vanishing of the first line implies the constraints
\begin{align}
H_{\a ab}=T_{\a(ab)}=F_{a\a I}+(\g_a)_{\a\b} W_I{}^\b = 0 .
\label{HTF}
\end{align}
Using the Bianchi identity involving $R_{[a\a\b]}{}^\g$ in the second line, we obtain the combination
\begin{align}
2\psi^a \L^\a (\g_a\L)_\b w_\g \left( 2Y_\a{}^{\b\g} + G_\a{}^{\b\g} \right) + \psi^a (\L\g^b\L) w_\a \left( C_{[ab]}{}^\a - G_{ab}{}^\a - T_{ab}{}^\a \right) ,
\label{cb}
\end{align}
which vanishes when
\begin{align}
G_{ab}{}^\a = C_{[ab]}{}^\a - T_{ab}{}^\a,\quad G_\a{}^{\b\g} = - 2Y_\a{}^{\b\g} .
\label{x5x6}
\end{align}
Using this, the last line of ${\cal G}_1$ becomes
\begin{align}
\L^\a \psi^a \psi^b \psi^c \left( \frac13 \N_\a T_{abc} - \N_a T_{\a bc} + T_{\a a}{}^d T_{dbc} - T_{\a\a}{}^\b T_{\b cb} - (\g_a)_{\a\b} T_{bc}{}^\b - R_{\a abc} \right) .
\label{lastG1}
\end{align}
We now prove that this expression vanishes. First we note that the Bianchi identity involving $\N_{[a} H_{b\a\b]}$ together with the constraint (\ref{cc5}) implies that
\begin{equation}
    H_{abc}+T_{abc}=0.
\end{equation}
Then, the Bianchi identity involving $\N_{[\a} H_{abc]}$ leads to
\begin{align}
\N_\a T_{abc} = -T_{\a[a}{}^d T_{bc]d} - (\g_{[a})_{\a\b} T_{bc]}{}^\b .
\label{NTabc}
\end{align}
Equation (\ref{lastG1}) then becomes
\begin{align}
-\L^\a \psi^a \psi^b \psi^c \left( R_{\a abc} + 2(\g_a)_{\a\b} T_{bc}{}^\b + T_{a\a}{}^\b T_{\b bc} + \N_a T_{\a bc} \right) .
\label{g1eq4}
\end{align}
Finally, using
\begin{align}
\psi^a \psi^b \psi^c R_{\a abc} = \frac12 \psi^a \psi^b \psi^c R_{\a [ab]c} =  \frac12 \psi^a \psi^b \psi^c \left( \N_{[\a} T_{ab]c} + T_{[\a a}{}^A T_{Ab]c} \right) ,
\label{Rabc}
\end{align}
and (\ref{NTabc}), we find that (\ref{lastG1}) vanishes. 

Up to now, we found that ${\cal G}_4={\cal G}_3={\cal G}_2={\cal G}_1=0$ when the background superfields satisfy certain constraints. Now take the contribution with zero $U(1)$-charge. Using the previous results, ${\cal G}_0$ is given by
\begin{align}
&{\cal G}_0 = 2 \left( \frac12 E_a E^a + d_\a E^a + \frac12 \psi_a\N\psi^a  + \frac12 w_\a \N\L^\a - \frac12 \L^\a \N w_\a \right) - w_\a \Nb \L^\a + \L^\a \Nb w_\a - \psi^a \Nb \psi_a \cr
&\qquad-2\L^\a w_\b \Jb^I \left( \N_\a W_I{}^\b + W_I{}^\g T_{\g\a}{}^\b \right) + \psi^a \psi^b \Jb^I \left( F_{abI} + W_I{}^\g T_{\g ab} \right)  \cr
&\qquad+2 \L^\a w_\b \B{E}^a T_{a\a}{}^\b + \psi^a \psi^b (\g_a\L)_\a w_\b \left( G_b{}^{\a\b} - 2Y_b{}^{\a\b} \right) \cr
&\qquad+\frac{1}{2}(\L\g^a\L) w_\a w_\b \left( G_a{}^{\a\b} - 2Y_a{}^{\a\b} \right) + \L^\a \L^\b w_\g w_\d T_{a\a}{}^\g T_{b\b}{}^{\d} \eta^{ab} .
\label{G0}
\end{align}
The first term in the right-hand side of this expression will help to obtain $-2T$. The vanishing of the third line implies the torsion constraint
\begin{align}
T_{a\a}{}^\b=  0 ,
\label{Taalpha}
\end{align}
and consequently
\begin{equation} \label{Omegaa}
    \Omega_a=\Omega_a\,^{bcde}=0,\qquad G_{\alpha b}\,^{\gamma}=C_{\alpha b}\,^{\gamma}
\end{equation}
as well as the constraint for
\begin{align}
G_a{}^{\a\b}= 2Y_a{}^{\a\b} .
\label{x7is}
\end{align}
Note that the last line in (\ref{G0}) is zero after using (\ref{Taalpha}) and (\ref{x7is}). We use the equations of motion for $\Lambda^{\alpha},w_{\alpha}$ and $\psi^a$ in (\ref{G0}) to obtain
\begin{align}
&{\cal G}_0 = 2 \left( \frac12 E_a E^a + d_\a E^a + \frac12 \psi_a\N\psi^a  + \frac12 w_\a \N\L^\a - \frac12 \L^\a \N w_\a \right) \cr
&-2\L^\a w_\b \Jb^I \left( -U_{I\a}{}^\b + \N_\a W_I{}^\b + W_I{}^\g T_{\g\a}{}^\b \right) + \psi^a \psi^b \Jb^I \left( U_{Iab} + F_{abI} + W_I{}^\g T_{\g ab} \right)  \cr
&+2\psi^a w_\a \left( \B{E}^A C_{Aa}{}^\a + \Jb^I U_{Ia}{}^\a \right) + 2w_\a w_\b \left( \B{E}^A Y_A{}^{\a\b} + \Jb^I Y_I{}^{\a\b} \right) .
\label{G00}
\end{align}
The vanishing of the second line leads to the constraints
\begin{align}
U_{I\a}{}^\b =\N_\a W_I{}^\b + W_I{}^\g T_{\g\a}{}^\b,\quad U_{Iab} = -\left(  F_{abI} + W_I{}^\g T_{\a ab} \right) .
\label{Uare}
\end{align}
Note that the second equation is a consequence of the Bianchi identity involving $\N_{[a} F_{I\a\b]}$.  

The last line in (\ref{G00}) has terms of charge $-1$ and $-2$, which will be added to ${\cal G}_{-1}$ and  ${\cal G}_{-2}$. Start with ${\cal G}_{-1}$, that is given by 
\begin{align}
{\cal G}_{-1} = 2\psi^a w_\a  E^A C_{Aa}{}^\a + 2 \psi^a w_\a \B{E}^b T_{ab}{}^\a + 2 \psi^a w_\a \Jb^I \N_a W_I{}^\a + \psi^a (\g_a\L)_\a w_\b w_\g G^{\a\b\g} .
\label{G-1}
\end{align}
The vanishing of the last term implies that $G^{\a\b\g}$ vanishes. The first term will contribute to finding $-2T$. 

Consider now ${\cal G}_{-2}$. It is given by
\begin{align}
&{\cal G}_{-2} = 2 w_\a w_\b E^A Y_A{}^{\a\b} + w_\a w_\b \Jb^I f_I\,^{JK} W_J{}^\a W_K{}^\b \cr
&\qquad\quad-\psi^a \psi^b w_\a w_\b \left( C_{ac}{}^\a C_{bd}{}^\b -  T_{ac}{}^\a G_{bd}{}^\b \right) \eta^{cd} .
\label{G-2}
\end{align}
The first term contributes to $-2T$. We now add $\m{G}_0,\m{G}_{-1}$ and $\m{G}_{-2}$. Cancellation of the terms with $ \B{E}^a$ and $\Jb^I$ implies the constraints
\begin{align}
&C_{ab}{}^\a = T_{ab}{}^\a = G_{ab}{}^\a,\quad C_{Ia}{}^\a =W_I\,^{\beta}C_{\beta a}\,^{\alpha} - \N_a W_I{}^\a,\cr
&Y_a\,^{\beta\gamma}=0,\quad Y_I{}^{\a\b} =W_I\,^{\gamma}Y_{\gamma}\,^{\alpha\beta}-\frac12 f_I\,^{JK} W_J{}^\a W_K{}^\b.
\label{CU}
\end{align}
Note that the first of these constraints implies that the second line of (\ref{G-2}) vanishes. The remaining contributions $\m{G}_{-3}$ and $\m{G}_{-4}$ vanish because they are proportional to either $Y_a\,^{\beta\gamma}$ or $G^{\alpha\beta\gamma}$. Finally, from the remaining terms in 
\begin{equation}
    \left( \oint  G\right)  G = \sum_{n=-4}^{4} {\cal G}_n,
\end{equation}
we obtain
\begin{align}
\left( \oint  G\right)  G= &-2  \Bigg( -\frac12E_aE^a-d_\a E^\a - \frac12 \psi_a \N \psi^a -\frac12 w_\a\N\L^\a +  \frac12 \L^\a \N w_\a\cr 
&\qquad\qquad\qquad- \psi^a w_\a E^A C_{Aa}{}^\a - w_\a w_\b E^A Y_A{}^{\a\b}   \Bigg)=-2T . \cr
\label{GGfinal}
\end{align}

\subsection{$\big\{G,G\big\}=-2T$ computation summary}
We now summarize the results of this section. We found that $G$ is given by
\begin{equation} \label{GF}
\begin{split}
    G=&\Lambda^{\alpha}d_{\alpha}+\psi^aE_a+w_{\alpha}E^{\alpha}\\
    &+\frac{1}{2}\big(\Lambda\gamma^a\Lambda\big)\psi_a-\frac{1}{2}\Lambda^{\alpha}\Lambda^{\beta}w_{\gamma}T_{\alpha\beta}\,^{\gamma}+\frac{1}{2}\Lambda^{\alpha}\psi^b\psi^cT_{\alpha bc}\\
    &+\Lambda^{\alpha}\psi^bw_{\gamma}C_{\alpha b}\,^{\gamma}-\frac{1}{6}\psi^a\psi^b\psi^cT_{abc}+\frac{1}{2}\psi^a\psi^bw_{\gamma}T_{ab}\,^{\gamma}-\Lambda^{\alpha}w_{\beta}w_{\gamma}Y_{\alpha}\,^{\beta\gamma}
\end{split}
\end{equation}
and that it squares to the stress-energy tensor when the background fields satisfy the constraints
\begin{align}
&T_{\alpha\beta}\,^{\gamma}=-G^{1440}_{\alpha\beta}\,^{\gamma}=-\widetilde{F}^{1440}_{abcd\delta}\big(\gamma_e\big)\,^{\delta\gamma}\big(\gamma^{abcde}\big)_{\alpha\beta},\cr 
&T_{\a\b a}=H_{\a\b a}=-(\g_a)_{\a\b} ,\quad T_{abc}=-H_{abc},\cr
&H_{\a\b\g}=F_{\a\b I}=T_{a\a}{}^\b=H_{\a ab}=T_{\a(ab)}=0,\cr
&F_{a\a I}+(\g_a)_{\a\b} W_I{}^\b = 0,\quad U_{I\a}{}^\b = \N_\a W_I{}^\b + W_I{}^\g T_{\g\a}{}^\b,\quad U_{Iab} = -  F_{abI} - W_I{}^\g T_{\g ab} ,\cr
&C_{Ia}{}^\a = W_I{}^\b C_{\b a}{}^\a - \N_a W_I{}^\a,\quad Y_I{}^{\a\b} = W_I{}^\g Y_\g{}^{\a\b} - \frac12 f_I\,^{JK} W_J{}^\a W_K{}^\b.
\label{rest}
\end{align}
It is important to stress that the torsions are defined with the $\big\{\Omega_{\alpha},\Omega_{\alpha}\,^{bc},\Omega_{a},\Omega_{a}\,^{bc},\Omega_{a}\,^{bcde}\big\}$ spin-connections. However, since $\Omega_{a}\,^{bcde}=0$ by (\ref{Omegaa}), the torsions reduce to those of \cite{Berkovits:2001ue}.

\section{Holomorphicity of $G$} \label{SectionHol}
We will now show that the constraints (\ref{rest}) imply that $G$ of (\ref{GF}) is holomorphic. The computation has the general form
\begin{equation}
    \B{\partial}G=\sum_{n=-3}^{+2}\big(\B{\partial}\m{G}\big)_n.
\end{equation}
In general, for each $\big(\B{\partial}\m{G}\big)_n$ there are two types of term, in a similar fashion as in the computation of the previous section. These terms are either Bianchi identities or are proportional to the constraints of (\ref{rest}). In both cases, it is clear that they vanish. We have compiled the relevant steps of the computation in Appendix \ref{apb}.

\subsection{Holomorphicity computation}
The natural next step after computing $\B{\partial}G$ using the information from Appendix \ref{apb} is the same as it was for $\big\{ G, G\big\}$: gather the contributions according to their factors and analyse the resulting conditions. We present the results in descending order of $U(1)$-charge.

The only contribution of $U(1)$-charge +2 comes from $\B{\partial}\Big(\frac{1}{2}(\Lambda\gamma^a\Lambda)\psi_a\Big)$. It is
\begin{equation}
\begin{split}
    \Lambda^{\alpha}\Lambda^{\beta}\psi^a\Big(&(\gamma_a)_{\alpha\beta}\B{ E}^A\Omega_A+(\gamma_a)_{\rho\alpha}\B{ E}^f\Omega_{fbcde}(\gamma^{bcde})_{\beta}\,^{\rho}\\
    &+(\gamma_a)_{\alpha\beta}\B{J}^IU_I+(\gamma_a)_{\rho\alpha}\B{J}^IU_{Ibcde}(\gamma^{bcde})_{\beta}\,^{\rho}\Big).
\end{split}
\end{equation}
After using $\B{ E}^{\alpha}=-\B{J}^IW_I\,^{\alpha}$, we conclude that this contribution vanishes if:
\begin{gather} \label{hol-1.1}
    (\gamma_a)_{\alpha\beta}\Omega_{f}+{\Omega}_{f}\,^{bcde}(\gamma_{abcde})_{\alpha\beta}=0,
\end{gather}
\begin{equation}\label{hol-1.2}
    U_I\delta^{\alpha}_{\beta}+{U}_{I}\,^{abcd}(\gamma_{abcd})_{\alpha}\,^{\beta}=W_I\,^{\rho}\Omega_{\rho}\delta_{\alpha}^{\beta}.
\end{equation}
The constraint (\ref{hol-1.1}) was derived in the previous section, whereas (\ref{hol-1.2}) can be shown to be equivalent to the Bianchi identity involving the super Yang-Mills field strength components $\nabla_{[\alpha}F_{\beta a]I}$.

For $U(1)$-charge +1, we also show in detail how the constraints (\ref{rest}) imply the vanishing of the contributions. For the simpler cases in the upcoming charges, however, we just mention the constraints that are involved, as the mechanism is essentially the same and the related factors can be recovered by the index structure and by requiring the correct conformal weight and $U(1)$-charge. The vanishing of the contribution
\begin{equation}
    \Lambda^{\alpha} E^a\B{J}^I\Big(F_{a\alpha I}+\frac{1}{2}W_I\,^{\rho}H_{\rho a\alpha}-\frac{1}{2}W_I\,^{\rho}T_{\rho\alpha a}\Big)
\end{equation}
is implied by the previously found constraints
\begin{equation}
    F_{a\alpha I}=-W_I\,^{\beta}(\gamma_a)_{\beta\alpha},\quad T_{\alpha\beta a}=H_{\alpha\beta a}=-(\gamma_a)_{\alpha\beta}.
\end{equation}
The vanishing of the contribution
\begin{equation}
    \Lambda^{\alpha} E^{\beta}\B{J}^I\Big(F_{\alpha\beta I}+\frac{1}{2}W_I\,^{\gamma}H_{\gamma\beta\alpha}\Big),
\end{equation}
on the other hand, is implied by the constraints
\begin{equation}
    F_{\a\b I}=-\frac{1}{2}W_I\,^{\gamma}H_{\gamma\alpha\beta}=0.
\end{equation}
Also, the vanishing of the contribution
\begin{equation}
    \Lambda^{\alpha}\B{J}^I\Big(\nabla_{\alpha}W_I\,^{\beta}+W_I\,^{\gamma}T_{\gamma\alpha}\,^{\beta}-U_{I\alpha}\,^{\beta}\Big)d_{\beta}
\end{equation}
is a consequence of the constraint
\begin{equation}
    U_{I\alpha}\,^{\beta}=\nabla_{\alpha}W_I\,^{\beta}+W_I\,^{\gamma}T_{\gamma\alpha}\,^{\beta}.
\end{equation}
Similarly, the vanishing of
\begin{equation}
    \Lambda^{\alpha} E^a\B{ E}^b\Big(T_{a\alpha b}+T_{b\alpha a}-H_{ab\alpha}\Big)
\end{equation}
is implied by the constraints
\begin{equation}
    H_{\alpha ab}=T_{\alpha(ab)}=0.
\end{equation}
Finally, the contribution
\begin{equation}
    \frac{1}{2}\Lambda^{\alpha} E^{\beta}\B{ E}^a\Big(T_{\alpha\beta a}-H_{a\beta\alpha}\Big)
\end{equation}
vanishes because of the constraint
\begin{equation}
    T_{\alpha\beta a}=H_{\alpha\beta a}.
\end{equation}

Now we move on to the Bianchi identity contributions. The term
\begin{equation}
    \frac{1}{2}\Lambda^{\alpha}\Lambda^{\beta}w_{\gamma}\B{ E}^a\Big(R_{a(\alpha\beta)}\,^{\gamma}-\nabla_{a}T_{\alpha\beta}\,^{\gamma}-(\gamma^b)_{\alpha\beta}T_{ab}\,^{\gamma}\Big)
\end{equation}
is just the Bianchi identity involving $R_{[a\alpha\beta]}\,^{\gamma}$ after we recall the constraint $T_{\alpha\beta}\,^a=-(\gamma^a)_{\alpha\beta}$, and therefore vanishes. The next term is
\begin{equation}
    \frac{1}{2}\Lambda^{\alpha}\psi^a\psi^b\B{ E}^c\Big(R_{\alpha cab}+\nabla_cT_{\alpha ab}+2(\gamma_a)_{\alpha\beta}T_{bc}\,^{\beta}\Big).
\end{equation}
After using the fact that
\begin{equation}
    2R_{\alpha cab}=R_{\alpha [ca]b}+R_{\alpha[bc]a}-R_{\alpha [ab]c}
\end{equation}
and the Bianchi identities involving $R_{[\alpha ca]b}, R_{[\alpha bc]a}$ and $R_{[\alpha ab]c}$, it can be shown to be equivalent to the Bianchi identity related to $\nabla_{[\alpha}H_{abc]}$. As a consequence, it also vanishes. Next, take the term
\begin{equation}
\begin{split}
    \frac{1}{2}\Lambda^{\alpha}\Lambda^{\beta}w_{\gamma}\B{J}^I\Big((\gamma^a)_{\alpha\beta}&\nabla_aW_I\,^{\gamma}+W_I\,^{\rho}\nabla_{\rho}T_{\alpha\beta}\,^{\gamma}+U_{I(\alpha}\,^{\rho}T_{\beta)\rho}\,^{\gamma}\\
    &-U_{I\rho}\,^{\gamma}T_{\alpha\beta}\,^{\rho}-\nabla_{(\alpha}U_{I\beta)}\,^{\gamma}-W_I\,^{\rho}R_{\rho(\alpha\beta)}\,^{\gamma}\Big).
\end{split}
\end{equation}
Once we use the constraints involving the $U_{I\alpha}\,^{\beta}$ field and the definition of the commutator $\big\{\nabla_{\alpha},\nabla_{\beta}\big\}W_I\,^{\rho}=(\gamma^a)_{\alpha\beta}\nabla_aW_I\,^{\rho}+W_I\,^{\rho}R_{\alpha\beta\rho}\,^{\gamma}$, said contribution can be seen to be equivalent to the Bianchi identity involving $R_{(\rho\alpha\beta)}\,^{\gamma}$. Finally, we have the term
\begin{equation}
\begin{split}
    \frac{1}{2}\Lambda^{\alpha}\psi^a\psi^b\B{J}^I\Big(W_I\,^{\rho}&R_{\rho\alpha ab}-\nabla_{\alpha}U_{Iab}-W_I\,^{\rho}\nabla_{\rho}T_{\alpha ab}\\
    &-U_{I\alpha}\,^{\rho}T_{\rho ab}+U_{Ic[a}T_{|\alpha| b]}\,^c+(\gamma_{[a})_{\alpha\beta}\nabla_{b]}W_I\,^{\beta}\Big).
\end{split}
\end{equation}
This contribution can be shown to be proportional to the Bianchi identity related to $R_{[\rho\alpha a]}\,^{\beta}$ after using the constraints involving the $U_I$ fields once again, as well as the Bianchi identity for $\nabla_{[\alpha}F_{ab]I}$.

The constraints that imply the vanishing of the contributions of zero U(1)-charge which are not combinations of Bianchi identities are the following:
\begin{gather}
    F_{a\alpha I}=-W_I\,^{\gamma}(\gamma_a)_{\gamma\alpha},\\
    U_{Iab}=-F_{abI}-W_I\,^{\rho}T_{\rho ab},\\
    C_{Ia}{}^\a = W_I{}^\b C_{\b a}{}^\a - \N_a W_I{}^\a,\\
    T_{abc}=-H_{abc},\\
    T_{\alpha(ab)}=H_{\alpha ab}=0.
\end{gather}
For the Bianchi identity contributions, start with
\begin{equation}
    \Lambda^{\alpha}w_{\beta}\psi^a\B{ E}^b\Big(R_{ab\alpha}\,^{\beta}-T_{ab}\,^{\rho}T_{\alpha\rho}\,^{\beta}-\nabla_{\alpha}T_{ab}\,^{\beta}-T_{b\alpha}\,^cT_{ca}\,^{\beta}-T_{b}\,^{c\beta}T_{\alpha ac}\Big)=0
\end{equation}
which is simply the Bianchi identity involving $R_{[ab\alpha]}\,^{\beta}$. Next, take
\begin{equation}
\frac{1}{2}\psi^a\psi^b\psi^c\B{ E}^d\Big(R_{dabc}-T_{da}\,^{\alpha}T_{\alpha bc}-\frac{1}{3}\nabla_dT_{abc}\Big)=0.
\end{equation}
This can be shown to reduce to the Bianchi identity involving $\nabla_{[d}H_{abc]}$ after we use the constraint $T_{abc}=-H_{abc}$ and the decomposition $R_{d[ab]c}=R_{[dab]c}+R_{abcd}$. Likewise, the term
\begin{equation}
\begin{split}
    \frac{1}{2}\psi^a\psi^b\psi^c\B{J}^I\Big(\nabla_aU_{Ibc}+\nabla_aW_I\,^{\rho}T_{\rho bc}+U_{Idc}T_{ab}\,^d+\frac{1}{3}W_I\,^{\rho}\nabla_{\rho}T_{abc}-W_I\,^{\rho}R_{\rho abc}\Big)
\end{split}
\end{equation}
can be shown to be zero after making use of the Bianchi identities involving $\nabla_{[a}F_{bc]I}$, $\nabla_{[\alpha}H_{abc]}$ and $R_{[\rho ab]c}$, as well as the constraints (\ref{rest}). Finally, after some manipulations with the constraint involving $C_{Ia}\,^{\beta}$ and the constraint involving $U_{I\alpha}\,^{\beta}$, the contribution that comes with a factor of $\Lambda^{\alpha}w_{\beta}\psi^a\B{J}^I$ can be written as
\begin{equation}
\begin{split}
    \Lambda^{\alpha}w_{\beta}\psi^a\B{J}^I\Big(W_I\,^{\rho}(&R_{\rho a\alpha}\,^{\beta}+(\gamma^b)_{\rho\alpha}T_{ab}\,^{\beta}+\nabla_aT_{\rho\alpha}\,^{\beta})\\
    &-T_{\alpha a}\,^b\nabla_bW_I\,^{\beta}+(\gamma_a)_{\alpha\rho}f_I\,^{JK}W_J\,^{\rho}W_K\,^{\beta}+[\nabla_a,\nabla_{\alpha}]W_I\,^{\beta}\Big).
\end{split}
\end{equation}
Now we use the definition of the commutator $[\nabla_a,\nabla_{\alpha}]W_I\,^{\rho}$ as well as the constraint $F_{\alpha aI}=W_I\,^{\rho}(\gamma_a)_{\alpha \rho}$. In doing so, we reduce this expression to the Bianchi identity involving $R_{[a\alpha\rho]}\,^{\beta}$.

The constraints:
\begin{gather}
    U_{I\alpha}\,^{\beta}=\nabla_{\alpha}W_I\,^{\beta}+W_I\,^{\gamma}T_{\gamma\alpha}\,^{\beta},\\
    C_{Ia}{}^\a = W_I{}^\b C_{\b a}{}^\a - \N_a W_I{}^\a,\\ 
    Y_I{}^{\a\b} = W_I{}^\g Y_\g{}^{\a\b} - \frac12 f_I\,^{JK} W_J{}^\a W_K{}^\b,
\end{gather}
imply the vanishing of the contributions with $-1$ U(1)-charge which are not Bianchi identities.
The first Bianchi contribution is
\begin{equation}
\begin{split}
    w_{\alpha}\psi^a\psi^b\B{ E}^c\Big(\nabla_aT_{cb}\,^{\alpha}-\nabla_cT_{ab}\,^{\alpha}+\frac{1}{2}T_{abd}T_c\,^{d\alpha}-T_{ca}\,^dT_{db}\,^{\alpha}-T_{ca}\,^{\rho}C_{\rho b}\,^{\alpha}-T_{cb}\,^{\rho}C_{\rho a}\,^{\alpha}\Big)=0
\end{split}
\end{equation}
and reduces to the Bianchi identity involving $\nabla_{[a}T_{bc]}\,^{\alpha}$. For the contribution 
\begin{equation}
\begin{split}
w_{\alpha}\psi^a\psi^b\B{J}^I\Big(\nabla_a&C_{Ib}\,^{\alpha}-\nabla_a\big(W_I\,^{\beta}C_{\beta b}\,^{\alpha}\big)-\frac{1}{2}f_I\,^{JK}(U_{Jab}+W_J\,^{\beta}T_{\beta ab})W_K\,^{\alpha}\\
&+W_I\,^{\rho}T_{\rho a}\,^cT_{cb}\,^{\alpha}+\frac{1}{2}W_I\,^{\rho}\nabla_{\rho}T_{ab}\,^{\alpha}-\frac{1}{2}T_{ab}\,^c\nabla_cW_I\,^{\alpha}-\frac{1}{2}U_{I\rho}\,^{\alpha}T_{ab}\,^{\rho}\Big)
\end{split}
\end{equation}
we use the constraint involving $\nabla_aW_I\,^{\alpha}$ such that the first two terms above become $-\nabla_a\nabla_bW_I\,^{\alpha}$. Then, by using the definition of the commutator:
\begin{equation}
    \big[\nabla_a,\nabla_b\big]W_I\,^{\alpha}=-T_{ab}\,^A\nabla_AW_I\,^{\alpha}+W_I\,^{\beta}R_{ab\beta}\,^{\alpha}+f_I\,^{JK}F_{ab J}W_K\,^{\alpha},
\end{equation}
it can be reduced to the Bianchi identity involving precisely $R_{[ab\beta]}\,^{\alpha}$. Finally,
\begin{equation}
\begin{split}
    -\Lambda^{\alpha}w_{\beta}w_{\gamma}\B{J}^I\big(\nabla_{\alpha}Y_I\,^{\beta\gamma}+W_I\,^{\rho}\nabla_{\alpha}&Y_\rho\,^{\beta\gamma}+W_I\,^{\sigma}T_{\sigma\alpha}\,^{\rho}Y_{\rho}\,^{\beta\gamma}-U_{I\alpha}\,^{\rho}Y_{\rho}\,^{\beta\gamma}\\
    &+f_I\,^{JK}U_{J\alpha}\,^{\beta}W_K\,^{\gamma}-f_I\,^{JK}W_J\,^{\rho}T_{\rho\alpha}\,^{\gamma}W_K\,^{\beta}\big)
\end{split}
\end{equation}
can be shown to be identically zero after using the constraints involving $\nabla_{\alpha}W_I\,^{\beta}$ and the $Y$ fields.

The contribution with $-2$ $U(1)$-charge is given by:
\begin{equation}
    w_{\alpha}w_{\beta}\psi^a\B{J}^I\big(\nabla_a(Y_I\,^{\alpha\beta}-W_I\,^{\rho}Y_{\rho}\,^{\alpha\beta})+f_I\,^{JK}\nabla_aW_J\,^{\alpha}W_K\,^{\beta}\big),
\end{equation}
where we have made use of the $C$ fields constraints from the previous section. It can be rewritten as
\begin{equation}
    f_I\,^{JK}\nabla_aW_J\,^{\alpha}W_K\,^{\beta}-f_I\,^{JK}\nabla_aW_J\,^{\alpha}W_K\,^{\beta}=0,
\end{equation}
after the use of the constraint involving the $Y$ fields.

Finally, the only non-trivial contribution with $-3$ $U(1)$-charge is given by:
\begin{equation}
    w_{\alpha}w_{\beta}w_{\gamma}\B{J}^If_I\,^{JK}\big(W_J\,^{\rho}Y_{\rho}\,^{\alpha\beta}W_K\,^{\gamma}-Y_I\,^{\alpha\beta}W_K\,^{\gamma}\big).
\end{equation}
After using the constraint involving the $Y$ fields, it can be shown that this expression is equivalent to
\begin{equation}
    f_I\,^{J[K}f_J\,^{LM]}W_L\,^{\alpha}W_M\,^{\beta}W_K\,^{\gamma},
\end{equation}
which is zero by the Jacobi identity $f_I\,^{J[K}f_J\,^{LM]}=0$ of the Yang-Mills algebra.

\section{Analysis of the constraints} \label{Analysis}
In the previous sections, we have shown that $\{G,G\}=-2T$ when
\begin{equation} \label{GG}
\begin{split}
    G=&\Lambda^{\alpha}d_{\alpha}+\psi^a E_a+w_{\alpha} E^{\alpha}\\
    &+\frac{1}{2}\big(\Lambda\gamma^a\Lambda\big)\psi_a-\frac{1}{2}\Lambda^{\alpha}\Lambda^{\beta}w_{\gamma}T_{\alpha\beta}\,^{\gamma}+\frac{1}{2}\Lambda^{\alpha}\psi^b\psi^cT_{\alpha bc}\\
    &+\Lambda^{\alpha}\psi^bw_{\gamma}C_{\alpha b}\,^{\gamma}-\frac{1}{6}\psi^a\psi^b\psi^cT_{abc}+\frac{1}{2}\psi^a\psi^bw_{\gamma}T_{ab}\,^{\gamma}-\Lambda^{\alpha}w_{\beta}w_{\gamma}Y_{\alpha}\,^{\beta\gamma}
\end{split}
\end{equation}
and the background fields satisfy the following set of constraints
\begin{equation} \label{CT0}
    T_{\alpha\beta}\,^{\gamma}=-\widetilde{F}^{1440}_{abcd\delta}\big(\gamma_e\big)\,^{\delta\gamma}\big(\gamma^{abcde}\big)_{\alpha\beta},
\end{equation}
\begin{equation} \label{C1}
    H_{\alpha\beta\gamma}=F_{\alpha\beta I}=T_{a\beta}\,^{\gamma}=H_{\alpha bc}=T_{\alpha(bc)}=0,
\end{equation}
\begin{equation}
    T_{\alpha\beta}\,^c=H_{\alpha\beta d}\eta^{dc}=-\big(\gamma^c\big)_{\alpha\beta},
\end{equation}
\begin{equation}
    T_{abc}=-H_{abc},
\end{equation}
\begin{equation}
    F_{a\beta I}=W_I\,^{\gamma}T_{\gamma\beta a},
\end{equation}
\begin{equation}
    U_{I\alpha}\,^{\beta}=\nabla_{\alpha}W_I\,^{\beta}+W_I\,^{\gamma}T_{\gamma\alpha}\,^{\beta},
\end{equation}
\begin{equation} \label{Uab}
    U_{Iab}=-F_{abI}-W_I\,^{\gamma}T_{\gamma ab},
\end{equation}
\begin{equation} \label{Cdet}
    C_{Ia}\,^{\beta}-W_I\,^{\gamma}C_{\gamma a}\,^{\beta}=-\nabla_aW_I\,^{\beta},
\end{equation}
\begin{equation} \label{Ydet}
    Y_I\,^{\alpha\beta}-W_I\,^{\gamma}Y_{\gamma}\,^{\alpha\beta}=-\frac{1}{2}f_I\,^{JK}W_J\,^{\alpha}W_K\,^{\beta}.
\end{equation}
As a consequence, the BRST charge is holomorphic.

Let us start our analysis of the constraints with (\ref{CT0}). The Bianchi identity
\begin{equation}
    \nabla_{(\alpha}T_{\beta\gamma)}\,^{d}+T_{(\alpha\beta}\,^ET_{|E|\gamma)}\,^d=0
\end{equation}
implies that $\widetilde{F}^{1440}_{abcd\delta}$ must be zero. In order to see why, note that it is not in the kernel of the linear map $T_{(\alpha\beta}\,^{\rho}\gamma^d_{\gamma)\rho}=0$ and that there are no other 1440 irrep contributions to this equation. Consequently, we conclude that $\widetilde{F}^{1440}_{abcd\delta}=0$ and also that
\begin{equation}
    T_{\alpha\beta}\,^{\gamma}=0.
\end{equation}
Then, this same Bianchi identity determines
\begin{equation}
    T_{\alpha b}\,^c=2\big(\gamma_b\,^c\big)_{\alpha}\,^{\beta}\Omega_{\beta}.
\end{equation}
The analysis of (\ref{C1}) to (\ref{Uab}) is straightforward, since they explicitly contain the rest of the super-Yang-Mills and supergravity constraints of \cite{Berkovits:2001ue}.

On the other hand, Equations (\ref{Cdet}) and (\ref{Ydet}) determine linear combinations of the $C$ and $Y$ background fields in terms of the super-Yang-Mills reduced field strength $W_I\,^{\alpha}$. 
These constraints can be simplified by gauging the $C_{\alpha b}\,^{\gamma}$ and $Y_{\alpha}\,^{\beta\gamma}$ background fields to zero using the following gauge symmetry of the BRST charge and worldsheet action:
\begin{equation}
    \delta d_{\alpha}=\psi^b w_{\gamma} \rho_{\alpha b}\,^{\gamma}-w_{\beta}w_{\gamma}\widetilde\rho_{\alpha}\,^{\beta\gamma},
\end{equation}
\begin{equation}
    \delta C_{Ia}\,^{\beta}=-W_I\,^{\gamma}\rho_{\gamma a}\,^{\beta},\qquad \delta Y_I\,^{\alpha\beta}=-W_I\,^{\gamma}\widetilde\rho_{\gamma}\,^{\alpha\beta},
\end{equation}
\begin{equation}
    \delta C_{\alpha b}\,^{\gamma}=-\rho_{\alpha b}\,^{\gamma},\qquad \delta Y_{\alpha}\,^{\beta\gamma}=- \widetilde\rho_{\alpha}\,^{\beta\gamma}.
\end{equation}
Then, the constraints (\ref{Cdet}) and (\ref{Ydet}) get rewritten as
\begin{equation}
    C_{Ia}\,^{\beta}=-\nabla_aW_I\,^{\beta},\qquad Y_I\,^{\alpha\beta}=-\frac{1}{2}f_I\,^{JK}W_J\,^{\alpha}W_K\,^{\beta}.
\end{equation}

It is also important to stress the fact that requiring $\big\{ G, G\big\}=-2T$ is not enough to recover all the constraints. In fact, it is easy to see that the U(1)-charge/small Hilbert space condition was paramount to reach this goal. Indeed, were we allowed to deform the $G$ current with terms in the large Hilbert space, we would lose both of the following essential constraints:
\begin{equation}
    T_{\alpha\beta}\,^c=-\big(\gamma^c\big)_{\alpha\beta},\qquad H_{\alpha\beta\gamma}=0.
\end{equation}

In conclusion, the action for the B-RNS-GSS heterotic string sigma-model in curved backgrounds is given by
\begin{equation} \label{Sfinal}
\begin{split}
S=\frac{1}{2 \pi\alpha'}\int d^2z\Bigg(&\frac{1}{2}E^a\B{E}_a+\frac{1}{2}E^B\B{E}^AB_{AB}+\frac{1}{2}\psi^a\B{\nabla}\psi_a+w_{\alpha}\B{\nabla}\Lambda^{\alpha}+d_{\alpha}\B{E}^{\alpha}+L_{gh}+L_{\B{J}}\\
&+E^AA_{AI}\B{J}^I+d_{\alpha}W_I\,^{\alpha}\B{J}^I-\frac{1}{2}\psi^a\psi^b\big(F_{abI}+W_I\,^{\gamma}T_{\gamma ab}\big)\B{J}^I+\Lambda^{\alpha}\nabla_{\alpha}W_I\,^{\beta}w_{\beta}\B{J}^I\\
&+\psi^b\B{E}^aT_{ab}\,^{\gamma}w_{\gamma}-\psi^a\nabla_aW_I\,^{\beta}w_{\beta}\B{J}^I-\frac{1}{2}w_{\alpha}w_{\beta}W_J\,^{\alpha}W_K\,^{\beta}f_I\,^{JK}\B{J}^I\Bigg),
\end{split}
\end{equation}
while the BRST charge is
\begin{align}
Q_B\equiv\oint d\s~ j_B = \oint d\s \Bigg( cT - c\Big(\frac{3}{2}\beta\partial\gamma+\frac{1}{2}\gamma\partial\beta\Big) + bc\p c + b\g^2 + \gamma G\Bigg),
\end{align}
with
\begin{equation} \label{Gfinal}
\begin{split}
    G=&\Lambda^{\alpha}d_{\alpha}+\psi^a E_a+w_{\alpha} E^{\alpha}\\
    &+\frac{1}{2}\big(\Lambda\gamma^a\Lambda\big)\psi_a+\frac{1}{2}\Lambda^{\alpha}\psi^b\psi^cT_{\alpha bc}\\
    &-\frac{1}{6}\psi^a\psi^b\psi^cT_{abc}+\frac{1}{2}\psi^a\psi^bw_{\gamma}T_{ab}\,^{\gamma}.
\end{split}
\end{equation}
As expected, this action includes the same $U(1)\times$Lorentz structure group as the action of \cite{Berkovits:2001ue}, the RNS variables $\psi^a$ and the deformations
\begin{equation}
    \psi^bw_{\gamma}\B{E}^aT_{ab}\,^{\gamma}-\psi^aw_{\beta}\B{J}^I\nabla_aW_I\,^{\beta}
\end{equation}
which are described by the B-RNS-GSS heterotic vertex of (\ref{SUGRAvertex}) and (\ref{SYMvertex}). However, it is surprising that the quadratic term
\begin{equation}
    -\frac{1}{2}w_{\alpha}w_{\beta}f_{I}\,^{JK}W_J\,^{\alpha}W_K\,^{\beta}\B{J}^I
\end{equation}
appears. Note that this new term does not appear in the vertex because it is a non-linear deformation.

\section{$\m{N}=(1,0)$ worldsheet superspace}

In this section, we reformulate the results from sections \ref{SectionG2}, \ref{SectionHol} and \ref{Analysis} with $\m{N}=(1,0)$ worldsheet superspace as the starting point. More precisely, by writing the most general $\m{N}=(1,0)$ worldsheet superconformal action in superspace, suitably introducing spin-connections and imposing the small Hilbert space restriction from Section \ref{SectionBRNSGSS}, we are able to derive the super-Yang-Mills and supergravity constraints. Consequently, we also verify that the action is $G$-exact.

We start by introducing the $\m{N}=(1,0)$ superspace coordinates $(z,\B{z},\kappa)$ and the superconformal generator
\begin{equation}
    \m{D}\equiv i\Big(\frac{\partial}{\partial\kappa}+\kappa\frac{\partial}{\partial z}\Big)
\end{equation}
such that $\m{D}^2=-\partial$. We also define the following $\m{N}=(1,0)$ superfields
\begin{equation} \label{SZ}
    \mathbb{Z}^M\equiv Z^M-i\kappa\Lambda^{\alpha}E_{\alpha}\,^M+i\kappa\psi^aE_a\,^M,
\end{equation}
\begin{equation} \label{SR}
    \B{\mathbb{R}}_{\m{A}}\equiv i\B{\rho}_{\m{A}}+\kappa \B{r}_{\m{A}},
\end{equation}
\begin{equation} \label{SW}
    \mathbb{W}_{\alpha}\equiv iw_{\alpha}+\kappa h_{\alpha}.
\end{equation}
We should clarify that $\B{\mathbb{R}}_{\m{A}}$ is the supermultiplet containing the heterotic fermions $\B{\rho}_{\m{A}}$ and their auxiliary variables $\B{r}_{\m{A}}$. We also recall the definition
\begin{equation}
    \B{J}^I\equiv\frac{1}{2}t^I_{\m{A}\m{B}}\B{\rho}_{\m{A}}\B{\rho}_{\m{B}},\quad t^I_{\m{A}\m{B}}\in\mathfrak{g}.
\end{equation}
Note that, up to sign conventions, the superfields (\ref{SZ}) to (\ref{SW}) are the curved background generalization of the $\m{N}=(1,0)$ superfields which appear in \cite{Berkovits:2016xnb}. The worldsheet supersymmetry transformations are given by
\begin{equation} \label{dG1}
    \delta_GZ^M=\Lambda^{\alpha}E_{\alpha}\,^M-\psi^aE_a\,^M,
\end{equation}
\begin{equation}
    \delta_G\psi^c=E^c-\frac{1}{2}\Lambda^{\alpha}\Lambda^{\beta}f_{\alpha\beta}\,^{c}+\frac{1}{2}\psi^b\psi^af_{ab}\,^c-\psi^b\Lambda^{\alpha}f_{\alpha b}\,^c,
\end{equation}
\begin{equation}
    \delta_G\Lambda^{\gamma}=-E^{\gamma}+\frac{1}{2}\Lambda^{\alpha}\Lambda^{\beta}f_{\alpha\beta}\,^{\gamma}-\frac{1}{2}\psi^b\psi^af_{ab}\,^{\gamma}-\Lambda^{\beta}\psi^af_{a\beta}\,^{\gamma},
\end{equation}
\begin{equation} \label{dW}
    \delta_Gw_{\alpha}=h_{\alpha},\qquad \delta_Gh_{\alpha}=-\partial w_{\alpha},
\end{equation}
\begin{equation} \label{dG4}
    \delta_G\B{\rho}_{\m{A}}=\B{r}_{\m{A}},\qquad\delta_G\B{r}_{\m{A}}=-\partial\B{\rho}_{\m{A}},
\end{equation}
where we recall the notation
\begin{equation}
    f_{MN}\,^A\equiv \partial_{[M}E_{N]}\,^A,\qquad f_{AB}\,^C=(-)^{A(N+B)}E_B\,^NE_A\,^Mf_{MN}\,^C.
\end{equation}

The most general super-Lagrangian one can write (ghost-number 0 and conformal weight (1/2,1)) is given by
\begin{equation}
\begin{split}
    \m{L}=-\frac{1}{2 \pi\alpha'}\Bigg(&\frac{i}{2}\Big(\eta_{ab}E_M\,^a\big(\mathbb{Z})E_N\,^b(\mathbb{Z})+{B}_{MN}(\mathbb{Z})\Big)\m{D}\mathbb{Z}^N\B{\partial}\mathbb{Z}^M-\B{\partial}\mathbb{Z}^M{E}_M\,^{\beta}(\mathbb{Z})\mathbb{W}_{\beta}\\
    &+\frac{i}{2}\B{\mathbb{R}}_{\m{A}}\m{D}\B{\mathbb{R}}_{\m{A}}+i\m{D}\mathbb{Z}^MA_M\,^{\m{A}\m{B}}(\mathbb{Z})\B{\mathbb{R}}_{\m{A}}\B{\mathbb{R}}_{\m{B}}+W_{\m{A}\m{B}}\,^{\alpha}(\mathbb{Z})\mathbb{W}_{\alpha}\B{\mathbb{R}}_{\m{A}}\B{\mathbb{R}}_{\m{B}}+\m{L}_{gh}\Bigg)
\end{split}
\end{equation}
where $\m{L}_{gh}=-i\mathbb{B}\B{\partial}\mathbb{C}$ denotes the super-Lagrangian for the ghost superfields $\mathbb{B}\equiv i\beta+\kappa b$ and $\mathbb{C}\equiv ic+\kappa\gamma$.
As usual, we obtain the ordinary Lagrangian by computing $L=\int d\kappa\m{L}=\m{D}\m{L}|_{\kappa=0}.$ After integrating out the auxiliary variable $\B{r}_{\m{A}}$ via its equation of motion 
\begin{equation}
    \B{r}_{\m{B}}=\Lambda^{\alpha}A_{\alpha\m{A}\m{B}}\B{\rho}_{\m{A}}-\psi^aA_{a\m{A}\m{B}}\B{\rho}_{\m{A}}-W_{\m{A}\m{B}}\,^{\alpha}w_{\alpha}\B{\rho}_{\m{A}}
\end{equation}
and using that $\B{\rho}_{\m{A}}\B{\rho}_{\m{B}}t^J_{\m{A}\m{B}}t^K_{\m{BC}}=f_I\,^{JK}\B{J}\,^I$,
we find that, up to total derivatives, the Lagrangian is given by:
\begin{equation} \label{N=1full}
\begin{split}
    L=\frac{1}{2\pi\alpha'}\Bigg(\frac{1}{2}&E^a\B{E}_a+\frac{1}{2}E^B\B{E}^AB_{AB}+h_{\alpha}\B{E}^{\alpha}+w_{\alpha}\B{\partial}\Lambda^{\alpha}+\frac{1}{2}\psi_a\B{\partial}\psi^a+L_{gh}+L_{\B{J}}\\
    &-\frac{1}{4}\Lambda^{\alpha}\Lambda^{\beta}\big(T_{\alpha\beta}\,^c-H_{\alpha\beta d}\eta^{dc}\big)\B{E}_c-\frac{1}{4}\Lambda^{\alpha}\Lambda^{\beta}H_{\alpha\beta\gamma}\B{E}^{\gamma}-\frac{1}{2}\Lambda^{\alpha}\Lambda^{\beta}F_{\alpha\beta I}\B{J}^I\\
    &+\frac{1}{2}\Lambda^{\alpha}\psi^b\big(H_{\alpha bc}-T_{\alpha(bc)}\big)\B{E}^c+\frac{1}{2}\Lambda^{\alpha}\psi_c\big(T_{\alpha\beta}\,^c+H_{\alpha\beta d}\eta^{dc}\big)\B{E}^{\beta}-\Lambda^{\alpha}\psi^bF_{\alpha bI}\B{J}^I\\
    &-\frac{1}{2}\B{E}^a\psi^b\psi^c\Big(\frac{1}{2}f_{bca}-f_{abc}-\frac{1}{2}H_{abc}\Big)-\frac{1}{2}\psi^a\psi^b\Big(f_{\gamma ab}+\frac{1}{2}H_{ab\gamma}\Big)\B{E}^{\gamma}-\frac{1}{2}\psi^a\psi^bF_{abI}\B{J}^I\\
    &-\B{E}^af_{a\beta}\,^{\gamma}\Lambda^{\beta}w_{\gamma}-\B{E}^{\alpha}f_{\alpha\beta}\,^{\gamma}\Lambda^{\beta}w_{\gamma}+f_{\alpha b}\,^{\gamma}\psi^bw_{\gamma}\B{E}^{\alpha}-\B{E}^aT_{ab}\,^{\gamma}\psi^bw_{\gamma}\\
    &+E^AA_{AI}\B{J}^I+h_{\alpha}W_I\,^{\alpha}\B{J}^I\\
    &+\Lambda^{\alpha}\nabla_{\alpha}W_I\,^{\beta}w_{\beta}\B{J}^I-\psi^a\nabla_aW_I\,^{\beta}w_{\beta}\B{J}^I-\frac{1}{2}w_{\alpha}w_{\beta}W_J\,^{\alpha}W_{K}\,^{\beta}f_{I}\,^{JK}\B{J}^I\Bigg).
\end{split}
\end{equation}
Note that here the covariant derivatives are defined solely with respect to the Yang-Mills gauge group, as we have not introduced spin connections yet. We stress the fact that the expressions above describe the most general superconformal Lagrangian that can be written with the untwisted variables $(\Lambda^{\alpha}$, $w_{\alpha})$. 

Now we introduce the spin-connections by choosing them to be given by
\begin{equation}
\Omega_{A\beta}\,^{\gamma}=\Omega_A\delta_{\beta}^{\gamma}+\frac{1}{4}\Omega_A\,^{bc}\big(\gamma_{bc}\big)_{\beta}\,^{\gamma},
\end{equation}
\begin{equation} \label{SCchoice1}
\Omega_{[ab]c}=-f_{abc}-H_{abc},\qquad \Omega_{a\beta}\,^{\gamma}=-f_{a}\delta_{\beta}^{\gamma}-\frac{1}{4}f_{abc}\big(\gamma^{bc}\big)_{\beta}\,^{\gamma},
\end{equation}
\begin{equation} \label{SCchoice2}
    \Omega_{(\alpha\beta)}\,^{\gamma}+f_{\alpha\beta}\,^{\gamma}=f^{1440}_{abcd\delta}\big(\gamma_e\big)\,^{\delta\gamma}\big(\gamma^{abcde}\big)_{\alpha\beta},
\end{equation}
where for the choice involving $\Omega_{(\alpha\beta)}\,^{\gamma}$ we recall the discussion that precedes Equation (\ref{red4}) at the beginning of Section \ref{SectionG2}.

The next step is to impose the U(1)-charge/small Hilbert space requirement on the deformations of the action. Before we do so, however, consider plugging the flat space values for the background superfields in (\ref{N=1full}). The expression reduces to
\begin{equation}
    \begin{split}
    S_{B-RNS-GSS}=\int d^2z\Big(\frac{1}{2}\Pi_m\overline \Pi^{m}+B_{het}+\frac{1}{2}\psi^m\B{\partial}\psi_m+h_{\alpha}\B{\partial}\theta^{\alpha}+w_{\alpha}\B{\partial}\Lambda^{\alpha}\\
    -\Lambda^{\alpha}\psi_m\gamma^m_{\alpha\beta}\B{\partial}\theta^{\beta}+b\B{\partial}c+\beta\B{\partial}\gamma+\B{b}\partial\B{c}+\frac{1}{2}\overline\rho^{\mathcal{A}}\partial\overline\rho_{\mathcal{A}}\Big).
    \end{split}
\end{equation}
Note that there is a positive U(1)-charge interaction term proportional to $\Lambda^{\alpha}\psi_m\gamma^m_{\alpha\beta}\B{\partial}\theta^{\beta}$ in this action, which is not present in the flat B-RNS-GSS action of (\ref{SBRNSGSS}). The presence of this term will cause problems when defining the notion of small Hilbert space, and consequently when imposing restrictions on the allowed deformations. Therefore, we should define $h_{\alpha}$ in terms of the other worldsheet variables as
\begin{equation}
    d_{\alpha}= h_{\alpha}-\Lambda^{\beta}\psi_m\gamma^m_{\alpha\beta}
\end{equation}
in order to take this action into the free expression of (\ref{SBRNSGSS}). 

Now we go back to the general expression (\ref{N=1full}) and carry over this last discussion. We conclude that we must similarly define
\begin{equation} \label{shiftpositive}
    d_{\alpha}=h_{\alpha}+\frac{1}{2}\Lambda^{\beta}\psi^c\big(T_{\alpha\beta c}+H_{\alpha\beta c}\big)-\frac{1}{4}\Lambda^{\beta}\Lambda^{\gamma}H_{\alpha\beta\gamma}-\frac{1}{2}\psi^b\psi^cT_{\alpha bc}-\Lambda^{\beta}w_{\gamma}\Omega_{\beta\alpha}\,^{\gamma}+\psi^b\Omega_{b\alpha}\,^{\gamma}w_{\gamma}
\end{equation}
before we apply the U(1)-condition from Section \ref{SectionBRNSGSS}, where it is convenient to also incorporate terms with zero and negative U(1)-charge in order to obtain a covariant expression for the action as in (\ref{Sfinal}). Indeed, as a result, we obtain covariant derivatives acting on $W_I\,^{\alpha},\psi^a$ and $\Lambda^{\alpha}$ which now also include the Lorentz and scaling connections. Note that, after defining $d_{\alpha}$, Equation (\ref{dW}) is amended to
\begin{equation}\label{conv}
    \delta_Gw_{\alpha}=d_{\alpha}-\frac{1}{2}\Lambda^{\beta}\psi^c\big(T_{\alpha\beta c}+H_{\alpha\beta c}\big)+\frac{1}{4}\Lambda^{\beta}\Lambda^{\gamma}H_{\alpha\beta\gamma}+\frac{1}{2}\psi^b\psi^cT_{\alpha bc}+\Lambda^{\beta}w_{\gamma}\Omega_{\beta\alpha}\,^{\gamma}-\psi^b\Omega_{b\alpha}\,^{\gamma}w_{\gamma}.
\end{equation}

Now we are able to impose the U(1)-charge/small Hilbert space condition, that is, require that deformations of the action and of $G$ lie in the small Hilbert space with respect to the untwisted variables ($\Lambda^{\alpha},w_{\alpha}$). From (\ref{N=1full}) and (\ref{conv}), 
we conclude that
\begin{equation} \label{CS1}
    T_{\alpha\beta}\,^c=H_{\alpha\beta d}\eta^{dc}=-\big(\gamma^c\big)_{\alpha\beta},\quad F_{a\beta I}=-\big(\gamma_a\big)_{\beta\gamma}W_I\,^{\gamma},
\end{equation}
\begin{equation} \label{CS2}
    H_{\alpha\beta\gamma}=0,\quad F_{\alpha\beta I}=-\frac{1}{2}W_I\,^{\gamma}H_{\gamma\alpha\beta}=0,
\end{equation}
\begin{equation} \label{CS3}
    H_{\alpha bc}=T_{\alpha(bc)}=0.
\end{equation}
In order to see how we get the rest of the constraints, recall that
\begin{equation} \label{TorsionS7}
    T_{AB}\,^C=f_{AB}\,^C+\Omega_{[AB]}\,^C.
\end{equation}
Then, the choice of spin-connections in (\ref{SCchoice1}) and (\ref{SCchoice2}) together with the constraints (\ref{CS1}) to (\ref{CS3}) imply that the constraints
\begin{equation}
T_{abc}=-H_{abc},\qquad T_{a\beta}\,^{\gamma}=0,\qquad T_{\alpha\beta}\,^{\gamma}=0,\qquad T_{\alpha bc}=2\big(\gamma_{bc}\big)_{\alpha}\,^{\beta}\Omega_{\beta},
\end{equation}
are satisfied. The first constraint can be verified simply from the definition (\ref{TorsionS7}). In order to verify the second one, we also need to use the Bianchi identity involving $\nabla_{[\alpha}H_{\beta cd]}$, as well as the Bianchi identity involving $\nabla_{[\alpha}T_{\beta c]}\,^d$. The last two constraints follow from the Bianchi identity involving $\nabla_{(\alpha}T_{\beta\gamma)}\,^d$, as discussed in the previous section.

Finally, the action becomes
\begin{equation}
\begin{split}
S=\frac{1}{2 \pi\alpha'}\int d^2z\Bigg(&\frac{1}{2}E^a\B{E}_a+\frac{1}{2}E^B\B{E}^AB_{AB}+\frac{1}{2}\psi^a\B{\nabla}\psi_a+w_{\alpha}\B{\nabla}\Lambda^{\alpha}+d_{\alpha}\B{E}^{\alpha}+L_{gh}+L_{\B{J}}\\
&+E^AA_{AI}\B{J}^I+d_{\alpha}W_I\,^{\alpha}\B{J}^I-\frac{1}{2}\psi^a\psi^b\big(F_{abI}+W_I\,^{\gamma}T_{\gamma ab}\big)\B{J}^I+\Lambda^{\alpha}\nabla_{\alpha}W_I\,^{\beta}w_{\beta}\B{J}^I\\
&+\psi^b\B{E}^aT_{ab}\,^{\gamma}w_{\gamma}-\psi^a\nabla_aW_I\,^{\beta}w_{\beta}\B{J}^I-\frac{1}{2}w_{\alpha}w_{\beta}W_J\,^{\alpha}W_K\,^{\beta}f_I\,^{JK}\B{J}^I\Bigg).
\end{split}
\end{equation}
Note that this is the same action as (\ref{Sfinal}). We have thus found the remaining constraints
\begin{equation}
U_{Iab}=-F_{abI}- W_I\,^{\gamma}T_{\gamma ab},\qquad U_{I\alpha}\,^{\beta}=\nabla_{\alpha}W_I\,^{\beta},
\end{equation}
\begin{equation}
 C_{Ia}\,^{\beta}=-\nabla_aW_I\,^{\beta},\qquad C_{ab}\,^{\gamma}=T_{ab}\,^{\gamma},\qquad Y_I\,^{\alpha\beta}=-\frac{1}{2}f_I\,^{JK}W_J\,^{\alpha}W_K\,^{\beta}\B{J}^I,
\end{equation}
recalling our initial notation set in (\ref{Sdyn}). Also, note that after the introduction of $d_{\alpha}$ and of the spin-connections, the worldsheet supersymmetry transformations on the worldsheet fields become
\begin{equation}
    \delta_G\psi^c=E^c+\frac{1}{2}\Lambda^{\alpha}\Lambda^{\beta}\gamma^c_{\alpha\beta}-\psi^b\psi^a\Omega_{ab}\,^c-\frac{1}{2}\psi^a\psi^bT_{ab}\,^c-\psi^b\Lambda^{\alpha}T_{\alpha b}\,^c+\psi^b\Lambda^{\alpha}\Omega_{\alpha b}\,^c,
\end{equation}
\begin{equation}
    \delta_G\Lambda^{\gamma}=-E^{\gamma}-\Lambda^{\alpha}\Lambda^{\beta}\Omega_{\alpha\beta}\,^{\gamma}-\frac{1}{2}\psi^b\psi^aT_{ab}\,^{\gamma}+\Lambda^{\beta}\psi^a\Omega_{a\beta}\,^{\gamma},
\end{equation}
\begin{equation}
    \delta_Gw_{\alpha}=d_{\alpha}+\gamma^c_{\alpha\beta}\psi_c\Lambda^{\beta}+\frac{1}{2}\psi^b\psi^cT_{\alpha bc}+\Lambda^{\beta}w_{\gamma}\Omega_{\beta\alpha}\,^{\gamma}-\psi^b\Omega_{b\alpha}\,^{\gamma}w_{\gamma},
\end{equation}
which allow us to recover $G$ of (\ref{Gfinal}).

We have thus derived the D=10 N=1 SUGRA/SYM constraints simply from writing the most general $\m{N}=(1,0)$ worldsheet superconformal action in superspace and correctly imposing the U(1)-charge/small Hilbert space constraint. Consequently, we have also verified that the action is $G$-exact.

\acknowledgments
NB would like to thank CNPq grant 311434/2020-7 and FAPESP grants 2016/01343-7, 2021/14335-0, 2019/21281-4 and 2019/24277-8 for partial financial support. OC would like to thank FONDECYT grants 1200342 and 1201550 for partial financial support. JG would like to thank FAPESP grants 2022/04105-0 and 2019/14061-8 for partial financial support. LNSM would like to thank FAPESP grants 2021/09003-9 and 2018/07834-8 for partial financial support.

\begin{appendices}
\section{Supergeometry review} \label{SuperGeo}
In this appendix we provide a review of supergeometry and some of the conventions used in the paper. See the standard reference \cite{Wess:1992cp} for more details. The supervielbein one-form is defined as $E^A=dZ^M E_M{}^A$, the supergravity connection one-form is defined as $\Omega_A{}^B=dZ^M \Omega_{MA}{}^B$ and the YM connection is defined as $A=A_I t^I=dZ^M A_{MI} t^I$, where $t^I$ are the generators of the Lie algebra of the heterotic gauge group. These generators satisfy the Lie algebra $[t^I,t^J]=f_K\,^{IJ} t^K$. The covariant derivative one-form acts on a $p$ form $\Psi^A=\Psi^A{}_I t^I$ as
\begin{align}
    \N\Psi^A=d\Psi^A+\Psi^B \Omega_B{}^A - \Psi^A A + (-1)^p A \Psi^A ,
    \label{COV}
\end{align}
where the product between forms is the wedge product. Acting with $\N$ we obtain the curvature and the field-strength 
\begin{align}
    \N\N\Psi^A=\Psi^B R_B{}^A - \Psi^A F + F \Psi^A , 
    \label{NN}
\end{align}
where 
\begin{align}
    R_A{}^B=d\Omega_A{}^B+\Omega_A{}^C\Omega_C{}^B,\quad F=dA-AA .
    \label{RandF}
\end{align}
The torsion two-form is defined as $T^A=\N E^A$. The two-forms $T^A, R_A{}^B, F$ satisfy Bianchi identities given by 
\begin{align}
    \N T^A = E^B R_B{}^A,\quad \N R_A{}^B=0,\quad dF=0 .
    \label{bianchis}
\end{align}
They can be written in components as
\begin{align}
    &\N_{[A} T_{BC]}{}^D + T_{[AB}{}^E T_{|E|C]}{}^A - R_{[ABC]}{}^D = 0 ,\cr
    &\N_{[A} R_{BC]D}{}^E + T_{[AB}{}^F R_{|F|C]D}{}^E = 0 ,\cr
    &\N_{[A} F_{BC]I} + T_{[AB}{}^D F_{|D|C]I} = 0 ,
    \label{bianchiCOMP}
\end{align}
where $T^A=\frac12 E^B E^C T_{CB}{}^A, R_B{}^A=\frac12 E^C E^D R_{DCB}{}^A, F=\frac12 E^B E^A F_{ABI} t^I$. The Kalb-Ramond field $B_{BA}$ in the action (\ref{Sdyn}) defines a two-form $B=\frac12 E^A E^B B_{BA}$ and its exterior derivative is $H=dB$ which satisfies the Bianchi identity $dH=0$. This identity can be written in components as
\begin{align}
    \N_{[A} H_{BCD]} + \frac32 T_{[AB}{}^E H_{|E|CD]}=0.
    \label{DH}
\end{align}

\section{Canonical commutators} \label{apa}

The canonical commutators for the variables in $G$ can be calculated from the basic canonical commutators (\ref{com1}), (\ref{com2}), (\ref{com3}) and (\ref{PZ}). We collect the results below.

\begin{align}
[ E_a(\s),f(Z(\s'))]=-\p_a f~\d(\s-\s') .
\label{}
\end{align}

\begin{align}
[d_\a(\s),f(Z(\s'))]=\p_\a f~\d(\s-\s') .
\label{}
\end{align}

\begin{align}
[ E_a(\s), E_b(\s')] =& \left( E_a{}^M(\s) E_{Mb}(\s') + E_a{}^M(\s) B_{Mb}(\s') \right) \frac{\p}{\p\s'} \d(\s-\s') \cr
&-\left( E_b{}^M(\s') E_{Ma}(\s) + E_b{}^M(\s') B_{Ma}(\s) \right) \frac{\p}{\p\s} \d(\s-\s') \cr
&+[ \p_\s Z^M \p_{[a} E_{Mb]} + \p_\s Z^N E_{[a}{}^M \p_{b]} B_{MN} - w_\a w_\b C_{ac}{}^\a C_{bd}{}^\b \eta^{cd} + R_{ab\a}{}^\b \L^\a w_\b \cr
& -\frac12 R_{abcd}\psi^c\psi^d + F_{abI}\Jb^I - T_{ab}{}^\a d_\a  +\p_\tau Z^M E_M{}^c T_{abc} - \p_\tau Z^M E_M{}^c\O_{[ab]c} \cr
&+\psi^c w_\a \left( \N_{[a} C_{b]c}{}^\a + T_{ab}{}^A C_{Ac}{}^\a\right)  + w_\a w_\b \left(\N_{[a} Y_{b]}{}^{\a\b} + T_{ab}{}^A Y_A{}^{\a\b} \right)] \d(\s-\s') \cr
\label{}
\end{align}

\begin{align}
[ E_a(\s) ,  E^\a(\s') ] =& \left( \p_a W_I{}^\a + f_I\,^{JK} A_{Ja} W_K{}^\a \right) \Jb^I \d(\s-\s')  + 2\p_\s Z^M \p_a E_M{}^\a \d(\s-\s')  \cr
&+2E_a{}^M(\s) E_M{}^\a(\s') \frac{\p}{\p\s'} \d(\s-\s')  .
\label{}
\end{align}

\begin{align}
[ E^\a(\s), E^\b(\s')]=f_I\,^{JK} W_J{}^\a W_K{}^\b \Jb^I \d(\s-\s') .
\label{}
\end{align}

\begin{align}
[d_\a(\s),d_\b(\s')]=&E_\a{}^M(\s) B_{M\b}(\s') \frac{\p}{\p\s'} \d(\s-\s')  + E_\b{}^M(\s') B_{M\a}(\s) \frac{\p}{\p\s} \d(\s-\s') \cr
&+[  w_\g w_\r C_{\a a}{}^\g C_{\b b}{}^\r \eta^{ab} + \p_\tau Z^M E_M{}^a T_{\a\b a} - T_{\a\b}{}^\g d_\g + \O_{(\a\b)}{}^\g d_\g \cr
&+ (-1)^N \p_\s Z^M E_{(\a}{}^N \p_{\b)} B_{NM} -\frac12 R_{\a\b ab}\psi^a \psi^b + R_{\a\b\g}{}^\r \L^\g w_\r + F_{\a\b I} \Jb^I \cr
&+\psi^a w_\g \left( \N_{(\a} C_{\b)a}{}^\g + T_{\a\b}{}^A C_{Aa}{}^\g \right) + w_\g w_\r \left( \N_{(\a} Y_{\b)}{}^{\g\r} + T_{\a\b}{}^A Y_A{}^{\g\r} \right) ]  \d(\s-\s')  \cr
\label{}
\end{align}

\begin{align}
[d_\a(\s), E_a(\s')]=&-\left( E_\a{}^M(\s) E_{Ma}(\s') + E_\a{}^M(\s) B_{Ma}(\s') \right)  \frac{\p}{\p\s'} \d(\s-\s') + E_a{}^M(\s') B_{M\a}(\s)  \frac{\p}{\p\s} \d(\s-\s') \cr
&+[ w_\b w_\g C_{\a b}{}^\b C_{ac}{}^\g \eta^{bc} +(-1)^{M+N} \p_\s Z^N E_a{}^M \p_\a B_{MN} - (-1)^N \p_\s Z^N E_\a{}^M \p_a B_{MN} \cr
&- (-1)^M \p_\s Z^M \p_\a E_{Ma} + \frac12 R_{\a abc}\psi^b\psi^c + R_{a\a\b}{}^\g \L^\b w_\g  + F_{a\a I} \Jb^I \cr 
&- \p_\tau Z^M E_M{}^b T_{\a ab} + \p_\tau Z^M E_M{}^b \O_{\a ab} \cr
&-T_{a\a}{}^\b d_\b + \O_{a\a}{}^\b d_\b + \psi^b w_\b \left( \N_{[\a} C_{a]b}{}^\b + T_{\a a}{}^A C_{Ab}{}^\b \right) ] \d(\s-\s') \cr
&- w_\b w_\g \left( \N_{[\a} Y_{a]}{}^{\b\g} + T_{\a a}{}^A Y_A{}^{\b\g} \right) \d(\s-\s') .
\label{}
\end{align}

\begin{align}
[d_\a(\s), E^\b(\s')]=&-2E_\a{}^M(\s) E_M{}^\b(\s') \frac{\p}{\p\s'} \d(\s-\s') -2(-1)^N \p_\s Z^N \p_\a E_N{}^\b \d(\s-\s') \cr
&-\left( \p_\a W_I{}^\b + f_I\,^{JK} A_{J\a} W_K{}^\b \right) \Jb^I \d(\s-\s') .
\label{}
\end{align}

\begin{align}
&[d_\a(\s),\L^\b(\s')]= \left( - \O_{\a\g}{}^\b \L^\g + \psi^a C_{\a a}{}^\b -2 Y_\a{}^{\b\g}w_\g \right) \d(\s-\s') .
\label{}
\end{align}

\begin{align}
&[d_\a(\s),w_\b(\s')]=\O_{\a\b}{}^\g w_\g \d(\s-\s') .
\label{}
\end{align}

\begin{align}
&[d_\a(\s),\psi_a(\s')]= \left( \O_{\a ab}\psi^b- w_\b C_{\a a}{}^\b \right) \d(\s-\s') .
\label{}
\end{align}

\begin{align}
&[ E_a(\s),\L^\a(\s')]=\left( \O_{a\b}{}^\a \L^\b + \psi^b C_{ab}{}^\a+2 Y_a{}^{\a\b} w_\b \right) \d(\s-\s') .
\label{}
\end{align}

\begin{align}
&[ E_a(\s),w_\a(\s')]=-\O_{a\a}{}^\b w_\b \d(\s-\s') .
\label{}
\end{align}

\begin{align}
&[ E_a(\s),\psi_b(\s')]= \left( -\O_{abc}\psi^c+ w_\a C_{ab}{}^\a \right) \d(\s-\s') .
\label{}
\end{align}

\section{Equations of motion for the worldsheet fields}\label{apb}
The relevant equations of motion to prove the holomorphicity of $G$ are
\begin{equation} \label{eomgbc}
    \B{\partial}\beta=\B{\partial}\gamma=\B{\partial}b=\B{\partial}c=0
\end{equation}
\begin{equation} \label{eomLambda}
    \B{\nabla}\Lambda^{\alpha}=-\Lambda^{\beta}\B{J}^IU_{I\beta}\,^{\alpha}+\psi^a\B{ E}^bT_{ab}\,^{\alpha}+\psi^a\B{J}^I\nabla_aW_I\,^{\alpha}+w_{\beta}\B{J}^If_I\,^{JK}W_J\,^{\beta}W_{K}\,^{\alpha}
\end{equation}

\begin{equation} \label{eomPsi}
    \B{\nabla}\psi_a=-\psi^b\B{J}^IU_{Iab}+ w_{\alpha}\B{ E}^bT_{ab}\,^{\alpha}+ w_{\alpha}\B{J}^I\nabla_aW_I\,^{\alpha}
\end{equation}

\begin{equation} \label{eomW}
    \B{\nabla}w_{\alpha}=w_{\beta}\B{J}^IU_{I\alpha}\,^{\beta}
\end{equation}
\begin{equation} \label{eomJ}
    \nabla\B{J}^I=f_K\,^{JI}\B{J}^K\Big(\frac{1}{2}U_{Jab}\psi^a\psi^b+U_{J\alpha}\,^{\beta}\Lambda^{\alpha}w_{\beta}+\psi^aw_{\alpha}C_{Ja}\,^{\alpha}+w_{\alpha}w_{\beta}Y_J\,^{\alpha\beta}+d_{\alpha}W_J\,^{\alpha}\Big)
\end{equation}
Note that we have used the constraints from Equation (\ref{rest}) in order to write the equations above.

The equations involving $\B{\nabla}d_{\alpha}$ and $\B{\nabla} E_a$, as well as their derivations, are more involved. Start by varying the action with respect to the supercoordinates \cite{Berkovits:2001ue}. We find:
\begin{equation} \label{eomZP}
    \frac{1}{2}\eta_{ab}\Big(-\B{\partial} E^aE_P\,^ b-E_P\,^a\partial\B{ E}^b-\partial Z^M\partial_{[M}E_{P]}\,^a\B{ E}^b- E^a\partial Z^M\partial_{[M}E_{P]}\,^b\Big)+\frac{1}{2} E^A\B{ E}^BH_{BAP}    
\end{equation}
$$
    +E_P\,^{\alpha}\B{\partial}d_{\alpha}+\B{\partial}Z^M\partial_{[M}E_{P]}\,^{\alpha}d_{\alpha}-\partial Z^M\partial_{[M}A_{P]I}\B{J}^I
$$
$$
-\Omega_{P\alpha}\,^{\beta}\B{\partial}\big(\Lambda^{\alpha}w_{\beta}\big)-\Lambda^{\alpha}w_{\beta}\B{\partial}Z^M\partial_{[M}\Omega_{P]\alpha}\,^{\beta}-\frac{1}{2}\Omega_P\,^{ba}\B{\partial}\big(\psi_a\psi_b\big)-\frac{1}{2}\psi_a\psi_b\B{\partial}Z^M\partial_{[M}\Omega_{P]}\,^{ba}
$$
$$
+C_P\,^{a\alpha}\B{\partial}\big(\psi_aw_{\alpha}\big)+\B{\partial}Z^M\partial_{[M}C_{P]}\,^{a\alpha}\psi_aw_{\alpha}-Y_P\,^{\alpha\beta}\B{\partial}\big(w_{\alpha}w_{\beta}\big)-w_{\alpha}w_{\beta}\B{\partial}Z^M\partial_{[M}Y_{P]}\,^{\alpha\beta}
$$
$$
    -A_{PI}f_K\,^{JI}\Big(\partial Z^MA_{MJ}+\frac{1}{2}U_J\,^{ab}\psi_a\psi_b+U_{J\alpha}\,^{\beta}\Lambda^{\alpha}w_{\beta}+\psi_aw_{\alpha}C_J\,^{a\alpha}+w_{\alpha}w_{\beta}Y_J\,^{\alpha\beta}+d_{\beta}W_J\,^{\beta}\Big)\B{J}^K
$$
$$
    -\B{J}^I\partial_PW_I\,^{\alpha}d_{\alpha}+\frac{1}{2}\B{J}^I\partial_PU_I\,^{ab}\psi_a\psi_b+\B{J}^I\partial_PU_{I\alpha}\,^{\beta}\Lambda^{\alpha}w_{\beta}-\B{J}^I\partial_PC_I\,^{a\alpha}\psi_aw_{\alpha}+\B{J}^I\partial_PY_I\,^{\alpha\beta}w_{\alpha}w_{\beta}=0.
$$
Note that
\begin{equation}
    \eta_{ab}\partial Z^M\partial_{[M}E_{P]}\,^a\B{ E}^b=\eta_{ab} E^A\B{ E}^bT_{AP}\,^a+\eta_{ab}E^c\Omega_{Pc}\,^a\B{E}^b-\eta_{ab}E_P\,^cE^A\Omega_{Ac}\,^a\B{E}^{b}
\end{equation}
such that the first line of Equation \ref{eomZP} can be rewritten as
\begin{equation}
    -\frac{1}{2}\eta_{ab}E_P\,^a\Big(\nabla\B{ E}^b+\B{\nabla} E^b\Big)-\frac{1}{2}\Big( E^A\B{ E}^a+\B{ E}^A E^a\Big)T_{APa}+\frac{1}{2} E^A\B{ E}^BH_{BAP}.
\end{equation}
Also note that
\begin{equation}
    \B{\partial}Z^M\partial_{[M}E_{P]}\,^{\alpha}d_{\alpha}=\B{\partial}Z^MT_{MP}\,^{\alpha}d_{\alpha}-E_P\,^{\beta}\B{\partial}Z^M\Omega_{M\beta}\,^{\alpha}d_{\alpha}-\B{J}^I\Omega_{P\beta}\,^{\alpha}W_I\,^{\beta}d_{\alpha}
\end{equation}
such that the terms with $d_{\alpha}$ can be rewritten as
\begin{equation}
    \B{ E}^AT_{AP}\,^{\alpha}+E_P\,^{\alpha}\B{\nabla}d_{\alpha}-\B{J}^I\nabla_PW_I\,^{\alpha}d_{\alpha}.
\end{equation}
The rest of Equation \ref{eomZP} can also be rewritten in a nicer form by recalling the definition of the covariant derivatives. In all, we will work with:
\begin{equation} \label{XPf}
    -\frac{1}{2}\eta_{ab}E_P\,^a\big(\nabla\B{ E}^b+\B{\nabla} E^b\big)-\frac{1}{2}\big( E^A\B{ E}^a+\B{ E}^A E^a\big)T_{APa}+\frac{1}{2} E^A\B{ E}^BH_{BAP}
\end{equation}
$$
+\B{ E}^AT_{AP}\,^{\alpha}d_{\alpha}+E_P\,^{\alpha}\B{\nabla}d_{\alpha}-\B{J}^I\nabla_PW_I\,^{\alpha}d_{\alpha}
$$
$$
-\frac{1}{2}\B{ E}^AR_{AP}\,^{ba}\psi_a\psi_b-\B{ E}^AR_{AP\alpha}\,^{\beta}\Lambda^{\alpha}w_{\beta}- E^AF_{API}\B{J}^I
$$
$$
+\B{ E}^AE_A\,^M\nabla_{[M}C_{P]a}\,^{\alpha}\psi^aw_{\alpha}-\B{J}^IC_P\,^{b\alpha}U_{Iba}\psi^aw_{\alpha}+\B{J}^IC_{Pa}\,^{\beta}U_{I\beta}\,^{\alpha}\psi^aw_{\alpha}
$$
$$
-\B{ E}^AE_A\,^M\nabla_{[M}Y_{P]}\,^{\alpha\beta}w_{\alpha}w_{\beta}-C_P\,^{a\alpha}\B{ E}^AC_{Aa}\,^{\beta}w_{\alpha}w_{\beta}-\B{J}^IC_P\,^{a\alpha}C_{Ia}\,^{\beta}w_{\alpha}w_{\beta}-2w_{\alpha}w_{\beta}\B{J}^IY_P\,^{\alpha\rho}U_{I\rho}\,^{\beta}
$$
$$
+\frac{1}{2}\psi^a\psi^b\B{J}^I\nabla_PU_{Iab}+\Lambda^{\alpha}w_{\beta}\B{J}^I\nabla_PU_{I\alpha}\,^{\beta}-\B{J}^I\nabla_PC_{Ia}\,^{\alpha}\psi^aw_{\alpha}+w_{\alpha}w_{\beta}\B{J}^I\nabla_PY_{I}\,^{\alpha\beta}=0.
$$
The following relation will also prove useful throughout the computation:
\begin{equation} \label{nPI}
    \B{\nabla} E^A-\nabla\B{ E}^A= E^C\B{ E}^BT_{BC}\,^A.
\end{equation}

The first terms we calculate are the ones coming from
\begin{equation}
    \B{\partial}\Big(\Lambda^{\alpha}d_{\alpha}+\psi^aE_a+w_{\alpha}E^{\alpha}\Big).
\end{equation}
The easiest is $\B{\partial}\big(w_{\alpha} E^{\alpha}\big)$. Clearly we have:
\begin{equation}
    \B{\partial}\big(w_{\alpha}E^{\alpha}\big)=w_{\alpha}\B{\nabla}E^{\alpha}+w_{\beta}\B{J}^IU_{I\alpha}\,^{\beta}E^{\alpha}.
\end{equation}
Then, by Equation \ref{nPI}, we may rewrite it as
\begin{equation}
    \B{\partial}\big(w_{\alpha} E^{\alpha}\big)=w_{\alpha}\big(\nabla\B{ E}^{\alpha}+ E^C\B{ E}^BT_{BC}\,^{\alpha}\big)+w_{\beta}\B{J}^IU_{I\alpha}\,^{\beta} E^{\alpha}.
\end{equation}
Next, recall the equation of motion for $\B{J}^I$ as well as $\B{ E}^{\alpha}=-\B{J}^IW_I\,^{\alpha}$, such that the final result is:
\begin{equation}
\begin{split}
    \B{\partial}\big(w_{\alpha} E^{\alpha}\big)=&w_{\alpha} E^C\B{ E}^BT_{BC}\,^{\alpha}-w_{\alpha}\B{J}^I\nabla W_I\,^{\alpha}+w_{\beta}\B{J}^IU_{I\alpha}\,^{\beta} E^{\alpha}
\end{split}
\end{equation}
$$
-w_{\rho}\Big(d_{\alpha}W_J\,^{\alpha}+\frac{1}{2}\psi^a\psi^bU_{Jab}+\Lambda^{\alpha}w_{\beta}U_{J\alpha}\,^{\beta}+\psi^aw_{\alpha}C_{Ja}\,^{\alpha}+w_{\alpha}w_{\beta}Y_J\,^{\alpha\beta}\Big)f_K\,^{JI}\B{J}^KW_I\,^{\rho}
$$

Now we move on to $\B{\partial}\big(\Lambda^{\alpha}d_{\alpha}\big)$. Again, it is easy to see that:
\begin{equation}
\begin{split}
    \B{\partial}\big(\Lambda^{\alpha}d_{\alpha}\big)=&\Lambda^{\alpha}\B{\nabla}d_{\alpha}-\Lambda^{\alpha}\B{J}^IU_{I\alpha}\,^{\beta}d_{\beta}\\
    &+\psi^a\B{ E}^bT_{ab}\,^{\alpha}d_{\alpha}+\psi^a\B{J}^I\nabla_aW_I\,^{\alpha}d_{\alpha}\\
    &+w_{\alpha}\B{J}^If_I\,^{JK}W_J\,^{\alpha}W_K\,^{\beta}d_{\beta}
\end{split}
\end{equation}
after using the constraints from Section \ref{SectionG2}. The tricky piece here is precisely $\Lambda^{\alpha}\B{\nabla}d_{\alpha}$, which we can obtain from Equation \ref{XPf} after isolating the term $E_P\,^{\alpha}\B{\nabla}d_{\alpha}$ and inverting the vielbein. It is explicitly given by:
\begin{equation}
\begin{split}
    \Lambda^{\alpha}\B{\nabla}d_{\alpha}=&\frac{1}{2}\Lambda^{\alpha}\big( E^A\B{ E}^a+ E^a\B{ E}^A\big)T_{A\alpha a}-\frac{1}{2}\Lambda^{\alpha} E^A\B{ E}^BH_{BA\alpha}\\
    &+\Lambda^{\alpha}\B{J}^I\nabla_{\alpha}W_I\,^{\beta}d_{\beta}-\Lambda^{\alpha}\B{ E}^AT_{A\alpha}\,^{\beta}d_{\beta}\\
    &+\frac{1}{2}\Lambda^{\alpha}\psi^a\psi^b\B{ E}^cR_{\alpha cab}+\frac{1}{2}\Lambda^{\alpha}\psi^a\psi^b\B{J}^IW_I\,^{\rho}R_{\rho\alpha ab}\\
    &+\Lambda^{\alpha}\Lambda^{\beta}w_{\gamma}\B{ E}^aR_{a\alpha\beta}\,^{\gamma}-\Lambda^{\alpha}\Lambda^{\beta}w_{\gamma}\B{J}^IW_I\,^{\rho}R_{\rho\alpha\beta}\,^{\gamma}+ \Lambda^{\alpha}E^AF_{A\alpha I}\B{J}^I\\
    &-\frac{1}{2}\Lambda^{\alpha}\psi^a\psi^b\B{J}^I\nabla_{\alpha}U_{Iab}-\Lambda^{\alpha}\Lambda^{\beta}w_{\gamma}\B{J}^I\nabla_{\alpha}U_{I\beta}\,^{\gamma}+\Lambda^{\alpha}w_{\beta}\psi^a\B{J}^I\nabla_{\alpha}C_{Ia}\,^{\beta}-\Lambda^{\alpha}w_{\beta}w_{\gamma}\B{J}^I\nabla_{\alpha}Y_I\,^{\beta\gamma}\\
    &-\Lambda^{\alpha}w_{\beta}\psi^a\B{ E}^b\nabla_{b}C_{\alpha a}\,^{\beta}+\Lambda^{\alpha}w_{\beta}\psi^a\B{ E}^b\nabla_{\alpha}T_{ba}\,^{\beta}-\Lambda^{\alpha}w_{\beta}\psi^a\B{ E}^bT_{b\alpha}\,^cT_{ca}\,^{\beta}\\
    &+\Lambda^{\alpha}w_{\beta}\psi^a\B{J}^IW_I\,^{\rho}\nabla_{(\rho}C_{\alpha)a}\,^{\beta}+\Lambda^{\alpha}w_{\beta}\psi^a\B{J}^I(W_I\,^{\rho}T_{\rho \alpha}\,^cT_{ca}\,^{\beta}+W_I\,^{\rho}T_{\rho\alpha}\,^{\sigma}C_{\sigma a}\,^{\beta})\\
    &+\Lambda^{\alpha}w_{\beta}\psi^a\B{J}^IC_{\alpha}\,^{b\beta}U_{Iba}-\Lambda^{\alpha}w_{\beta}\psi^a\B{J}^IC_{\alpha a}\,^{\rho}U_{I\rho}\,^{\beta}\\
    &+\Lambda^{\alpha}w_{\beta}w_{\gamma}\B{ E}^a\nabla_aY_{\alpha}\,^{\beta\gamma}\\
    &-\Lambda^{\alpha}w_{\beta}w_{\gamma}\B{J}^IW_I\,^{\rho}\nabla_{(\rho}Y_{\alpha)}\,^{\beta\gamma}-\Lambda^{\alpha}w_{\beta}w_{\gamma}\B{J}^IW_I\,^{\rho}T_{\rho\alpha}\,^{\sigma}Y_{\sigma}\,^{\beta\gamma}\\
    &+\Lambda^{\alpha}w_{\beta}w_{\gamma}\B{ E}^aC_{\alpha}\,^{b\beta}T_{ab}\,^{\gamma}-\Lambda^{\alpha}w_{\beta}w_{\gamma}\B{J}^IC_{\alpha}\,^{a\beta}\nabla_aW_I\,^{\gamma}+2\Lambda^{\alpha}w_{\beta}w_{\gamma}\B{J}^IY_{\alpha}\,^{\rho\beta}U_{I\rho}\,^{\gamma}
\end{split}
\end{equation}

The case for $\B{\partial}\big(\psi^a E_a\big)$ is similar. A simple computation yields the result
\begin{equation}
\begin{split}
    \B{\partial}\big(\psi^a E_a\big)=&\psi^a\B{\nabla} E_a-\psi^a\B{J}^IU_{Iba} E^b\\
    &-w_{\alpha}\B{ E}^a E^bT_{ab}\,^{\alpha}+w_{\alpha}\B{J}^I\nabla_aW_I\,^{\alpha} E^a
\end{split}
\end{equation}
where, again, $\psi^a\B{\nabla} E_a$ can be found from Equation \ref{XPf} after using Equation \ref{nPI} and inverting the appropriate vielbein. Its explicit expression is
\begin{equation}
\begin{split}
    \psi^a\B{\nabla} E_a=&\frac{1}{2} \psi_aE^C\B{ E}^BT_{BC}\,^a-\frac{1}{2}\psi^a\Big( E^A\B{ E}^b+\B{ E}^A E^b\Big)T_{Aab}+\frac{1}{2}\psi^a E^A\B{ E}^BH_{BAa}\\
    &+\psi^a\B{ E}^AT_{Aa}\,^{\alpha}d_{\alpha}-\psi^a\B{J}^I\nabla_aW_I\,^{\alpha}d_{\alpha}-\psi^a E^AF_{AaI}\B{J}^I\\
    &+\frac{1}{2}\psi^a\psi^b\psi^c\B{ E}^dR_{dabc}-\frac{1}{2}\psi^a\psi^b\psi^c\B{J}^IW_I\,^{\rho}R_{\rho abc}-\Lambda^{\alpha}w_{\beta}\psi^a\B{ E}^bR_{ba\alpha}\,^{\beta}+\Lambda^{\alpha}w_{\beta}\psi^a\B{J}^IW_I\,^{\rho}R_{\rho a\alpha}\,^{\beta}\\
    &+\frac{1}{2}\psi^a\psi^b\psi^c\B{J}^I\nabla_aU_{Ibc}+\Lambda^{\alpha}w_{\beta}\psi^a\B{J}^I\nabla_aU_{I\alpha}\,^{\beta}+ w_{\alpha}\psi^a\psi^b\B{J}^I\nabla_aC_{Ib}\,^{\alpha}+w_{\alpha}w_{\beta}\psi^a\B{J}^I\nabla_aY_I\,^{\alpha\beta}\\
    &- w_{\alpha}\psi^a\psi^b\B{ E}^c\nabla_cT_{ab}\,^{\alpha}+ w_{\alpha}\psi^a\psi^b\B{ E}^c\nabla_aT_{cb}\,^{\alpha}- w_{\alpha}\psi^a\psi^b\B{ E}^cT_{ca}\,^{\rho}C_{\rho b}\,^{\alpha}- w_{\alpha}\psi^a\psi^b\B{ E}^cT_{ca}\,^dT_{db}\,^{\alpha}\\
    &+ w_{\alpha}\psi^a\psi^b\B{J}^IW_I\,^{\rho}\nabla_{\rho}T_{ab}\,^{\alpha}- w_{\alpha}\psi^a\psi^b\B{J}^IW_I\,^{\rho}\nabla_{a}C_{\rho b}\,^{\alpha}+ w_{\alpha}\psi^a\psi^b\B{J}^IW_I\,^{\rho}T_{\rho a}\,^cT_{cb}\,^{\alpha}\\
    &+ w_{\alpha}\psi^a\psi^b\B{J}^IT_a\,^{c\alpha}U_{Icb}- w_{\alpha}\psi^a\psi^b\B{J}^IT_{ab}\,^{\beta}U_{I\beta}\,^{\alpha}\\
    &-w_{\alpha}w_{\beta}\psi^a\B{J}^IW_I\,^{\rho}\nabla_aY_{\rho}\,^{\alpha\beta}-w_{\alpha}w_{\beta}\psi^a\B{ E}^bT_a\,^{c\alpha}T_{bc}\,^{\beta}+w_{\alpha}w_{\beta}\psi^a\B{J}^IT_a\,^{b\alpha}\nabla_bW_I\,^{\beta}\\
    &-w_{\alpha}w_{\beta}\psi^a\B{ E}^bT_{ba}\,^{\rho}Y_{\rho}\,^{\alpha\beta}.
\end{split}
\end{equation}

The next step is computing the contribution coming from the $\frac{1}{2}(\Lambda\gamma^a\Lambda)\psi_a$ term in $G$. It is given by:
\begin{equation}
\begin{split}
    \B{\partial}\Big(\frac{1}{2}(\Lambda\gamma^a\Lambda)\psi_a\Big)=&-\Lambda^{\alpha}\Lambda^{\beta}\psi^a\Big((\gamma_a)_{\alpha\beta}\B{ E}^A\Omega_A+(\gamma_a)_{\rho\alpha}\B{ E}^a\Omega_{abcde}(\gamma^{bcde})_{\beta}\,^{\rho}\\
    &\qquad\qquad\qquad\quad+(\gamma_a)_{\alpha\beta}\B{J}^IU_I+(\gamma_a)_{\rho\alpha}\B{J}^IU_{Ibcde}(\gamma^{bcde})_{\beta}\,^{\rho}\Big)\\
    &+\Lambda^{\alpha}\psi^a\psi^b(\gamma_a)_{\alpha\beta}\big(\B{ E}^cT_{bc}\,^{\beta}+\B{J}^I\nabla_bW_I\,^{\beta}\big)\\
    &+\frac{1}{2}\Lambda^{\alpha}\Lambda^{\beta}w_{\gamma}(\gamma^a)_{\alpha\beta}\big(\B{ E}^bT_{ab}\,^{\gamma}+\B{J}^I\nabla_aW_I\,^{\gamma}\big)\\
    &+\Lambda^{\alpha}w_{\gamma}\psi^a\B{J}^I(\gamma_a)_{\alpha\beta}f_I\,^{JK}W_J\,^{\beta}W_K\,^{\gamma}
\end{split}
\end{equation}

Finally, we compute the anti-holomorphic derivative of the remaining terms in $G$. The results are compiled below:
\begin{equation}
\begin{split}
    -\frac{1}{2}\B{\partial}(\Lambda^{\alpha}\Lambda^{\beta}w_{\gamma}T_{\alpha\beta}\,^{\gamma})=&-\frac{1}{2}\Lambda^{\alpha}\Lambda^{\beta}w_{\gamma}\B{ E}^a\nabla_aT_{\alpha\beta}\,^{\gamma}+\frac{1}{2}\Lambda^{\alpha}\Lambda^{\beta}w_{\gamma}\B{J}^IW_I\,^{\rho}\nabla_{\rho}T_{\alpha\beta}\,^{\gamma}\\
    &+\Lambda^{\alpha}\Lambda^{\beta}w_{\gamma}\B{J}^IU_{I\alpha}\,^{\rho}T_{\rho\beta}\,^{\gamma}-\frac{1}{2}\Lambda^{\alpha}\Lambda^{\beta}w_{\gamma}\B{J}^IU_{I\rho}\,^{\gamma}T_{\alpha\beta}\,^{\rho}\\
    &+\Lambda^{\alpha}w_{\beta}\psi^a\B{ E}^bT_{ba}\,^{\rho}T_{\alpha\rho}\,^{\gamma}-\Lambda^{\alpha}w_{\beta}\psi^a\B{J}^I\nabla_aW_I\,^{\rho}T_{\alpha\rho}\,^{\beta}\\
    &-\Lambda^{\alpha}w_{\beta}w_{\gamma}\B{J}^If_I\,^{JK}W_J\,^{\rho}W_K\,^{\beta}T_{\alpha\rho}\,^{\gamma}
\end{split}
\end{equation}

\begin{equation}
\begin{split}
    \frac{1}{2}\B{\partial}(\Lambda^{\alpha}\psi^a\psi^bT_{\alpha ab})=&\frac{1}{2}\Lambda^{\alpha}\psi^a\psi^b\B{ E}^c\nabla_cT_{\alpha ab}-\frac{1}{2}\Lambda^{\alpha}\psi^a\psi^b\B{J}^IW_I\,^{\rho}\nabla_{\rho}T_{\alpha ab}\\
    &-\frac{1}{2}\Lambda^{\alpha}\psi^a\psi^b\B{J}^IU_{I\alpha}\,^{\rho}T_{\rho ab}-\Lambda^{\alpha}\psi^a\psi^b\B{J}^IU_{Icb}T_{\alpha a}\,^c\\
    &-\frac{1}{2}\psi^a\psi^b\psi^c\B{ E}^dT_{da}\,^{\alpha}T_{\alpha bc}+\frac{1}{2}\psi^a\psi^b\psi^c\B{J}^I\nabla_aW_I\,^{\rho}T_{\rho bc}\\
    &+\Lambda^{\alpha}w_{\beta}\psi^a\B{ E}^bT_b\,^{c\beta}T_{\alpha ca}+\Lambda^{\alpha}w_{\beta}\psi^a\B{J}^I\nabla_cW_I\,^{\beta}T_{\alpha a}\,^c\\
    &+\frac{1}{2}w_{\alpha}\psi^a\psi^b\B{J}^If_I\,^{JK}W_J\,^{\alpha}W_K\,^{\beta}T_{\beta ab}
\end{split}
\end{equation}

\begin{equation}
\begin{split}
    \B{\partial}(\Lambda^{\alpha}w_{\beta}\psi^aC_{\alpha a}\,^{\beta})=&\Lambda^{\alpha}w_{\beta}\psi^a\B{ E}^b\nabla_bC_{\alpha a}\,^{\beta}-\Lambda^{\alpha}w_{\beta}\psi^a\B{J}^IW_I\,^{\rho}\nabla_{\rho}C_{\alpha a}\,^{\beta}\\
    &+\Lambda^{\alpha}w_{\beta}\psi^a\B{J}^IU_{I\rho}\,^{\beta}C_{\alpha a}\,^{\rho}-\Lambda^{\alpha}w_{\beta}\psi^a\B{J}^IU_{I\alpha}\,^{\rho}C_{\rho a}\,^{\beta}-\Lambda^{\alpha}w_{\beta}\psi^a\B{J}^IU_{Iba}C_{\alpha }\,^{b\beta}\\
    &-\Lambda^{\alpha}w_{\beta}w_{\gamma}\B{ E}^aT_{a}\,^{b\beta}C_{\alpha b}\,^{\gamma}+\Lambda^{\alpha}w_{\beta}w_{\gamma}\B{J}^I\nabla_aW_I\,^{\beta}C_{\alpha}\,^{a\gamma}\\
    &-w_{\alpha}\psi^a\psi^b\B{ E}^cT_{cb}\,^{\beta}C_{\beta a}\,^{\alpha}+w_{\alpha}\psi^a\psi^b\B{J}^I\nabla_bW_I\,^{\beta}C_{\beta a}\,^{\alpha}\\
    &+\frac{1}{2}w_{\alpha}w_{\beta}\psi^a\B{J}^If_I\,^{JK}W_J\,^{\rho}W_K\,^{\alpha}C_{\rho a}\,^{\beta}
\end{split}
\end{equation}

\begin{equation}
\begin{split}
    -\frac{1}{6}\B{\partial}(\psi^a\psi^b\psi^cT_{abc})=&-\frac{1}{6}\psi^a\psi^b\psi^c\B{ E}^d\nabla_dT_{abc}+\frac{1}{6}\psi^a\psi^b\psi^c\B{J}^IW_I\,^{\rho}\nabla_{\rho}T_{abc}\\
    &+\frac{1}{2}\psi^a\psi^b\psi^c\B{J}^IU_{Idc}T_{ab}\,^d\\
    &+\frac{1}{2}\psi^a\psi^bw_{\alpha}\B{ E}^{c}T_{c}\,^{d\alpha}T_{abd}-\frac{1}{2}\psi^a\psi^bw_{\alpha}\B{J}^I\nabla_cW_I\,^{\alpha}T_{ab}\,^c
\end{split}
\end{equation}

\begin{equation}
\begin{split}
    \frac{1}{2}\B{\partial}(\psi^a\psi^bw_{\alpha}T_{ab}\,^{\alpha})=&\frac{1}{2}w_{\alpha}\psi^a\psi^b\B{ E}^c\nabla_cT_{ab}\,^{\alpha}-\frac{1}{2}w_{\alpha}\psi^a\psi^b\B{J}^IW_I\,^{\rho}\nabla_{\rho}T_{ab}\,^{\alpha}\\
    &-w_{\alpha}\psi^a\psi^b\B{J}^IU_{Icb}T_{a}\,^{c\alpha}+\frac{1}{2}w_{\alpha}\psi^a\psi^b\B{J}^IU_{I\rho}\,^{\alpha}T_{ab}\,^{\rho}\\
    &+w_{\alpha}w_{\beta}\psi^a\B{ E}^bT_{b}\,^{c\alpha}T_{ca}\,^{\beta}-w_{\alpha}w_{\beta}\psi^a\B{J}^I\nabla^bW_I\,^{\alpha}T_{ba}\,^{\beta}
\end{split}
\end{equation}

\begin{equation}
\begin{split}
    -\B{\partial}(\Lambda^{\alpha}w_{\beta}w_{\gamma}Y_{\alpha}\,^{\beta\gamma})=&-\Lambda^{\alpha}w_{\beta}w_{\gamma}\B{ E}^a\nabla_aY_{\alpha}\,^{\beta\gamma}+\Lambda^{\alpha}w_{\beta}w_{\gamma}\B{J}^IW_I\,^{\rho}\nabla_{\rho}Y_{\alpha}\,^{\beta\gamma}\\
    &+\Lambda^{\alpha}w_{\beta}w_{\gamma}\B{J}^IU_{I\alpha}\,^{\rho}Y_{\rho}\,^{\beta\gamma}-2\Lambda^{\alpha}w_{\beta}w_{\gamma}\B{J}^IU_{I\rho}\,^{\beta}Y_{\alpha}\,^{\gamma\rho}\\
    &+w_{\alpha}w_{\beta}\psi^a\B{ E}^bT_{ba}\,^{\rho}Y_{\rho}\,^{\alpha\beta}-w_{\alpha}w_{\beta}\psi^a\B{J}^I\nabla_aW_I\,^{\rho}Y_{\rho}\,^{\alpha\beta}\\
    &-w_{\alpha}w_{\beta}w_{\gamma}\B{J}^If_I\,^{JK}W_J\,^{\rho}W_K\,^{\alpha}Y_{\rho}\,^{\beta\gamma}.
\end{split}
\end{equation}
\end{appendices}
\bibliography{main}

\begin{thebibliography}{10}
\ifx\href\asklfhas\newcommand{\href}[2]{#2}\fi
\ifx\arxivref\asklfhas\newcommand{\arxivref}[2]{\href{http://arxiv.org/abs/#1}{#2}}\fi
\ifx\doiref\asklfhas\newcommand{\doiref}[2]{\href{http://dx.doi.org/#1}{#2}}\fi
\parskip 0pt
\normalsize

\bibitem{Friedan:1985ge}
D.~Friedan, E.~J. Martinec \& S.~H. Shenker,
\textit{``{Conformal Invariance, Supersymmetry and String Theory}''},
\doiref{10.1016/0550-3213(86)90356-1}{Nucl.~Phys.~B \textbf{271}, 93
  (1986)\ignorespaces}\ignorespaces
\bibitem{Berkovits:2013eqa}
N.~Berkovits,
\textit{``{Covariant Map Between Ramond-Neveu-Schwarz and Pure Spinor
  Formalisms for the Superstring}''},
\doiref{10.1007/JHEP04(2014)024}{JHEP \textbf{1404}, 024
  (2014)\ignorespaces}\ignorespaces,
\normalsize{\texttt{\arxivref{1312.0845}{arXiv:1312.0845
  \![hep-th]}}}\ignorespaces
\bibitem{Berkovits:2016xnb}
N.~Berkovits,
\textit{``{Untwisting the pure spinor formalism to the RNS and twistor string
  in a flat and AdS$_{5} \times$ S$^{5}$ background}''},
\doiref{10.1007/JHEP06(2016)127}{JHEP \textbf{1606}, 127
  (2016)\ignorespaces}\ignorespaces,
\normalsize{\texttt{\arxivref{1604.04617}{arXiv:1604.04617
  \![hep-th]}}}\ignorespaces
\bibitem{Berkovits:2021xwh}
N.~Berkovits,
\textit{``{Manifest spacetime supersymmetry and the superstring}''},
\doiref{10.1007/JHEP10(2021)162}{JHEP \textbf{2110}, 162
  (2021)\ignorespaces}\ignorespaces,
\normalsize{\texttt{\arxivref{2106.04448}{arXiv:2106.04448
  \![hep-th]}}}\ignorespaces
\bibitem{Siegel:1985xj}
W.~Siegel,
\textit{``{Classical Superstring Mechanics}''},
\doiref{10.1016/0550-3213(86)90029-5}{Nucl.~Phys.~B \textbf{263}, 93
  (1986)\ignorespaces}\ignorespaces
\bibitem{Berkovits:2000fe}
N.~Berkovits,
\textit{``{Super Poincare covariant quantization of the superstring}''},
\doiref{10.1088/1126-6708/2000/04/018}{JHEP \textbf{0004}, 018
  (2000)\ignorespaces}\ignorespaces,
\normalsize{\texttt{\arxivref{hep-th/0001035}{hep-th/0001035}}}\ignorespaces
\bibitem{Berkovits:2001ue}
N.~Berkovits \& P.~S. Howe,
\textit{``{Ten-dimensional supergravity constraints from the pure spinor
  formalism for the superstring}''},
\doiref{10.1016/S0550-3213(02)00352-8}{Nucl.~Phys.~B \textbf{635}, 75
  (2002)\ignorespaces}\ignorespaces,
\normalsize{\texttt{\arxivref{hep-th/0112160}{hep-th/0112160}}}\ignorespaces
\bibitem{Berkovits:2001us}
N.~Berkovits,
\textit{``{Relating the RNS and pure spinor formalisms for the superstring}''},
\doiref{10.1088/1126-6708/2001/08/026}{JHEP \textbf{0108}, 026
  (2001)\ignorespaces}\ignorespaces,
\normalsize{\texttt{\arxivref{hep-th/0104247}{hep-th/0104247}}}\ignorespaces
\bibitem{Kroyter:2009rn}
M.~Kroyter,
\textit{``{Superstring field theory in the democratic picture}''},
\doiref{10.4310/ATMP.2011.v15.n3.a3}{Adv.~Theor.~Math.~Phys. \textbf{15}, 741
  (2011)\ignorespaces}\ignorespaces,
\normalsize{\texttt{\arxivref{0911.2962}{arXiv:0911.2962
  \![hep-th]}}}\ignorespaces
\bibitem{Berkovits:2022}
N.~Berkovits,
\textit{``{D=5 Holomorphic Chern-Simons and the Pure Spinor Superstring}''},
\normalsize{\texttt{\arxivref{2211.06731}{arXiv:2211.06731
  \![hep-th]}}}\ignorespaces
\bibitem{Witten:1985nt}
E.~Witten,
\textit{``{Twistor - Like Transform in Ten-Dimensions}''},
\doiref{10.1016/0550-3213(86)90090-8}{Nucl.~Phys.~B \textbf{266}, 245
  (1986)\ignorespaces}\ignorespaces
\bibitem{Wess:1992cp}
J.~Wess \& J.~Bagger,
\textit{``{Supersymmetry and supergravity}''},
Princeton University Press (1992)\ignorespaces,
Princeton, NJ, USA
\end{thebibliography}
\nocite{*}
\end{document}